\documentclass[12pt]{article} 
\usepackage{array,epsfig,fancyheadings,rotating}
\usepackage[]{hyperref}  
\usepackage{sectsty, secdot, float}
\usepackage{natbib}
\usepackage{subfigure}
\usepackage{amsmath}
\usepackage{amssymb}
\usepackage{amsfonts}
\usepackage{multirow}
\usepackage{tabularx}
\usepackage{amsthm}
\usepackage{booktabs}
\usepackage{subfigure}
\usepackage[normalem]{ulem}
\useunder{\uline}{\ul}{}
\usepackage{booktabs}
\usepackage[mathscr]{eucal}

\usepackage{centernot}
\usepackage{lipsum}
\usepackage{setspace}
\usepackage{dlfltxbcodetips,bbm,mathtools,subfigure,amsfonts,amsmath,amssymb,fancybox,graphicx,bm,latexsym}

\usepackage{tikz}

\usetikzlibrary{bayesnet}
\usetikzlibrary{fit,positioning}

\newcommand{\logit}{\mbox{logit}}

\usepackage[tikz]{bclogo}

\newcommand{\bt}{\begin{bclogo}[couleur={rgb:orange,0;yellow,0;white,1},arrondi=0.1,logo=\bcplume,ombre=true]}
\newcommand{\et}{\end{bclogo}\s}
\newcommand{\btt}{\begin{box}}
\newcommand{\ett}{\end{box}}

\newcommand{\btheorem}{\begin{bclogo}[couleur={rgb:orange,0;yellow,0;white,1},arrondi=0.1,logo=\bcplume,ombre=true]{Theorem}}
\newcommand{\ettheorem}{\end{bclogo}}

\newcommand{\bst}{\begin{bclogo}[couleur={rgb:orange,1;yellow,1;white,0.5},arrondi=0.1,logo=\bcpanchant]}
\newcommand{\est}{\end{bclogo}}

\newcommand{\benum}{\begin{enumerate}}
\newcommand{\eenum}{\end{enumerate}}

\newcommand{\bq}{\begin{quote}\em}
\newcommand{\eq}{\end{quote}}
\newcommand{\bbq}{\begin{quote}\bf\em}
\newcommand{\ebq}{\end{quote}}

\newcommand{\ind}{\msim\limits^{\mbox{\tiny ind}}}
\newcommand{\iid}{\msim\limits^{\mbox{\tiny iid}}}

\newcommand{\mR}{\mathbb{R}}

\newcommand{\one}{\mathbbm{1}}

\newcommand{\mbP}{\mathbb{P}}

\newcommand{\hide}[1]{}
\newcommand{\ba}{\begin{array}{llllllllll}}
\newcommand{\ea}{\end{array}}
\newcommand{\bea}{\begin{equation}\begin{array}{llllllllll}}
\newcommand{\eea}{\end{array}\end{equation}}
\newcommand{\be}{\begin{equation}\begin{array}{lllllllllllllllll}}
\newcommand{\beno}{\begin{equation}\begin{array}{lllllllllllll}\nonumber}
\newcommand{\ee}{\end{array}\end{equation}}
\newcommand{\bel}{\begin{equation}\begin{array}{lllllllllllll}\nonumber}
\newcommand{\eel}{\Box\end{array}\end{equation}}
\newcommand{\bi}{\begin{itemize}}
\newcommand{\ei}{\end{itemize}}
\newcommand{\ben}{\begin{enumerate}}
\newcommand{\een}{\end{enumerate}}

\newcommand{\dsum}{\displaystyle\sum\limits}

\newcommand{\dprod}{\displaystyle\prod\limits}

\newcommand{\s}{\vspace{0.25cm}}
\newcommand{\bx}{\bm{x}}

\newcommand{\mT}{\mathscr{T}}

\newcommand{\balpha}{\bm{\alpha}}

\newcommand{\bbeta}{\bm{\beta}}

\newcommand{\bSigma}{\mbox{\boldmath$\Sigma$}}

\newcommand{\msim}{\mathop{\rm \sim}}

\newcounter{comment}
\newenvironment{comment}[1][]{\refstepcounter{comment}\vspace{0.25cm}\par\medskip\noindent%
\textbf{Comment~\thecomment #1}:\vspace{0.25cm}\\ \rmfamily}{\medskip}
\newcommand{\bc}{\begin{comment}\em}
\newcommand{\ec}{\end{comment}}

\newcounter{ex}

\newcounter{counterexample}

\newcounter{proposition}
\newenvironment{proposition}[1][]{\refstepcounter{proposition}\par\medskip\indent%
\textbf{Proposition~\theproposition #1}. \rmfamily}{\medskip}

\newcounter{result}

\newcounter{ttproof}
\newcounter{pproof}
\newcounter{ccproof}
\newcounter{llproof}
\newcounter{com}

\newcounter{lproof}
\newcounter{assumption}

\author{ }

\date{}

\begin{document}

\renewcommand{\baselinestretch}{2}





\fontsize{12}{14pt plus.8pt minus .6pt}\selectfont \vspace{0.8pc}
\centerline{\large\bf A CONTINUOUS-TIME STOCHASTIC PROCESS FOR}
\vspace{2pt} 
\centerline{\large\bf HIGH-RESOLUTION NETWORK DATA IN SPORTS}
\vspace{.4cm} 
\centerline{Nicholas Grieshop$^1$, Yong Feng$^2$, Guanyu Hu$^3$ and Michael Schweinberger$^4$}
\vspace{.4cm} 
\centerline{\it 1 Department of Statistics, University of Missouri-Columbia}
\vspace{.4cm} 
\centerline{\it 2 Department of Economics, University of Missouri-Columbia}
\begin{center}
{\it 3 Department of Biostatistics and Data Science,
The University of Texas Health Science Center at Houston}
\end{center}
\centerline{\it 4 Department of Statistics, The Pennsylvania State University}
\vspace{.55cm} \fontsize{9}{11.5pt plus.8pt minus.6pt}\selectfont


\begin{quotation}
\noindent 
{\it Abstract:}
Technological advances have paved the way for collecting high-resolution network data in basketball, football, and other team-based sports. Such data consist of interactions among players of competing teams indexed by space and time. High-resolution network data are vital to understanding and predicting the performance of teams, because the performance of a team is more than the sum of the strengths of its individual players: Whether a collection of players forms a strong team depends on the strength of the individual players as well as the interactions among the players. We introduce a continuous-time stochastic process as a model of interactions among players of competing teams indexed by space and time, discuss basic properties of the continuous-time stochastic process, and learn the stochastic process from high-resolution network data by pursuing a Bayesian approach.
We present simulation results along with an application to Juventus Turin, Inter Milan, and other football clubs in the premier Italian soccer league.

\vspace{9pt}
\noindent {\it Key words and phrases:}
Continuous-time stochastic processes; 
Relational event data; 
Soccer games;
Spatio-temporal data;
Sport analytics.
\par
\end{quotation}\par

\def\thefigure{\arabic{figure}}
\def\thetable{\arabic{table}}

\renewcommand{\theequation}{\thesection.\arabic{equation}}

\fontsize{12}{14pt plus.8pt minus .6pt}\selectfont

\section{Introduction}

Sport analytics has witnessed a surge of interest in the statistics community \citep[see, e.g.,][]{albert2017handbook},
driven by technological advances that have paved the way for collecting high-resolution tracking data in basketball, football, and other team-based sports.

Traditional sport analytics has focused on predicting match outcomes based on summary statistics \citep{dixon1997modelling, karlis2003analysis,baio2010bayesian,cattelan2013dynamic}. 
In more recent times,
the advent of high-resolution tracking data 
has expanded the role of statistics in sport analytics \citep{albert2017handbook} and has enabled granular evaluations of players and teams \citep{cervone2014pointwise,franks2015characterizing, cervone2016multiresolution,wu2018modeling,yurko2019nflwar,hu2021cjs} along with in-game strategy evaluations \citep{fernandez2018wide,sandholtz2019measuring,nguyen2023here}. 
High-resolution tracking data fall into two categories: 
optical ball- and player-tracking data obtained from video footage collected by multiple cameras in sport arenas,
and data collected by wearable devices. 
Some recent papers have used high-resolution tracking data to evaluate the defensive strength of teams \citep{franks2015characterizing};
constructing a dictionary of play types \citep{miller2017possession};
assessing the expected value of ball possession in basketball \citep{cervone2016multiresolution,santos2022role};
and constructing deep generative models of spatio-temporal trajectory data \citep{santos2022role}. 

{As a case in point,
we focus on soccer---that is,
European football.
Soccer is a fast-paced sport that generates high-resolution network data in the form of ball-tracking data indexed by space and time.
The statistical analysis of high-resolution network data generated by soccer poses many challenges,
including---but not limited to---the following:
\bi
\item[1.] Scoring a goal in a soccer match is a rare event, and useful predictors are hard to come by: 
e.g.,
a soccer team may score 0, 1, or 2 goals during a typical match, 
and scoring a goal requires a sequence of complex interactions among players of two competing teams.
\item[2.] Soccer teams consist of more players and the interactions among the players are more complex than,
e.g.,
in basketball and other team-based sports.
The fact that soccer teams are larger than teams in many other team-based sports implies that the actions of players on the field need to be coordinated. 
To facilitate coordination,
each soccer team adopts a formation,
which assigns each player in the team to a specific position (e.g., goalkeeper, striker).
Two popular formations of soccer teams,
known as 4-4-2 and 3-5-2,
are shown in Figure \ref{fig:formation} in Supplement \ref{sec:plots}.
The chosen formation can affect the defensive and offensive strategies of a soccer team and can hence affect the outcome of a match.
In addition,
players may have different roles in different formations,
and the formations of teams may change during matches.
\hide{
\item[3.] Changes in the compositions of teams due to substitutions of players,
changes in the formations of soccer teams (e.g., from 4-4-2 to 3-5-2), 
and goals can affect the pace of a soccer match:
e.g.,
a team on track to winning a match may decrease its pace and play more defensively,
while its opponent may increase its pace and play more offensively to change the outcome of the match in its favor.
}
\item[3.] Soccer matches are zero-sum games:
One team's gain is another team's loss.
For example,
if the ball changes hands,
one team loses control of the ball while the other team gains control of the ball.
\ei
}

{We address the lack of a comprehensive statistical analysis of the network of interactions among soccer players by introducing a continuous-time stochastic process,
which helps shed light on
\bi
\item which player controls the ball and how long,
and how ball control depends on the player's attributes (including the player's position in the team's formation and the player's spatial position on the field, 
provided that the spatial positions of players on the field are known);
\item whether a change in ball control is a failure (i.e., the ball is lost to the opposing team) or a success (i.e., the ball remains within the team in control of the ball),
and how the probability of a failure or a success depends on attributes of players;
\item whether a team on track to winning a match decreases its pace and plays more defensively,
while its opponent increases its pace and plays more offensively to change the outcome of the match in its favor;
\item unobserved attributes of players that may affect ball control and interactions among players.
\ei
}
{\subsection{Comparison with non-network models of sport data}\label{sec:literature1}}

{In contrast to the literature on basketball and other team-based sports,
we do not focus on individual summaries,
such as the expected ball possession of individual players \citep[e.g.,][]{cervone2016multiresolution,santos2022role}.
Instead,
we focus on the network of interactions among players,
because the performance of a team is more than the sum of the strengths of its players.
In other words,
a collection of strong players may or may not form a strong soccer team:
Whether a collection of players forms a strong soccer team depends on the one hand on the strength of the individual players and on the other hand on how the players interact.

Some recent publications \citep[e.g.,][]{chacoma2020modeling,hirotsu2023soccer,narizuka2023validation} have studied soccer matches by using probabilistic models, 
but the mentioned publications focus on time-dependent motion processes and ignore the network of interactions among players.
By contrast,
the proposed stochastic modeling framework focuses on the network of interactions among players and helps incorporate the formations of soccer teams in addition to the spatial distances between players,
provided that the spatial positions of players are known.

Compared with the continuous-time within-play valuation models of American football in \citet{yurko2020going}, 
the proposed stochastic modeling framework focuses on the pace of soccer matches,
who is in control of the ball, 
whether a change in ball control is a failure or a success,
and who secures control of the ball,
rather than focusing on action evaluations. 
As a result,
the proposed stochastic modeling framework can provide a more comprehensive understanding of team work in soccer and other team-based sports than the existing literature.}

{\subsection{Comparison with discrete-time models of sport data}\label{sec:literature3}}

{State-space models and other discrete-time stochastic processes have been used as predictive models for National Football League (NFL) game scores and other team-based sports \citep[e.g.,][]{glickman2005state,shaw2019dynamic}. 
By contrast,
we focus on continuous-time Markov processes,
for at least two reasons.

First,
continuous-time Markov processes are natural models of real-world processes where events can occur at any time $t \in [0, +\infty)$,
including fast-paced soccer matches.

Second,
continuous-time Markov processes can be viewed as discrete-time Markov chains with the time gaps between transitions of the Markov chains filled with Exponential holding times \citep[see, e.g., Chapter 3 of][]{Nj97}.
In other words,
continuous-time Markov processes model when changes take place, 
and which changes take place.
Therefore,
continuous-time Markov processes help build richer models than discrete-time Markov processes.
For example,
in applications to soccer matches,
continuous-time Markov processes help shed light on:
\bi
\item {\it Clock:} 
When a change in ball control occurs,
and how a change depends on the attributes of the player in control of the ball.
\item {\it Transitions:} 
Who passes the ball to whom,
and how a change in ball control depends on the attributes of the players involved.
\ei}

\vspace{-1cm}

{\subsection{Comparison with relational event models}
}
\label{sec:literature2}

{The closest relatives of the proposed stochastic modeling framework may be relational event models \citep[e.g.,][]{Butts2008,Perry2010}.
That said,
there are important differences between relational event models and the proposed stochastic modeling framework:\begin{itemize}
\item[1.] {\bf Soccer matches revolve around the ball.}
A reasonable stochastic model of soccer matches needs to reflect the fact that soccer matches revolve around the ball:
e.g.,
at any given time $t$,
a single player is in control of the ball and can initiate a relational event (e.g., a pass),
and a stochastic model of soccer matches should reflect that.
By contrast,
relational event models assume that any actor can initiate a relational event at any time:
e.g.,
at any given time $t$,
any employee of a company can send an email to one or more other employees.
\item[2.] {\bf Soccer matches are zero-sum games:}
One team's gain is another team's loss.
As a result,
a reasonable stochastic model of soccer matches should distinguish between successful and unsuccessful relational events (e.g., passes),
which can affect the outcomes of a match.
By contrast,
relational event models are not concerned with zero-sum games and do not distinguish between successful and unsuccessful relational events:
e.g.,
email communications between the employees of a company are not zero-sum games, 
and the event that employee A sends an email to employee B does not necessarily result in a gain or a loss for employee A or employee B.
\item[3.] {\bf The formations of soccer teams and the locations of players on the field can affect the outcome of a match.}
By contrast,
if an employee of a company considers sending an email,
the location of the employee is unimportant:
As long as the employee is connected to the World Wide Web,
the employee can send an email from any location on planet Earth.
\end{itemize}
}
\subsection{Structure of paper}
\label{sec:structure}

We first introduce the data that motivated the proposed stochastic modeling framework (Section \ref{sec:data}) and then introduce the stochastic modeling framework (Section \ref{sec:model}).
A Bayesian approach to learning the stochastic modeling framework from data is described in Section \ref{sec:inference}, 
and Bayesian computing is discussed in Section \ref{sec:computing}.
An application to the motivating data is presented in Section \ref{sec:app}.
Simulation results can be found in Section \ref{sec:simulations}.

\section{{High-resolution network data}}
\label{sec:data}

We consider data provided by Hudl \& Wyscout (\url{https://footballdata.wyscout.com/}).
The data consists of 380 matches during the 2020/21 season of Serie A, 
the premier league of the Italian football league system.
{The data include ball-tracking data,
but not player-tracking data.
In other words,
we know which player is in control of the ball,
but we do not know where the players are located on the field.}

Figure \ref{fig:network-comb} in Supplement \ref{sec:plots} shows a subset of the data: 
passes between the players of Juventus Turin (with 4-4-2 formation) and Inter Milan (with 3-5-2 formation).
These data are based on the home games of Juventus Turin versus AC Milan and Inter Milan versus AC Milan in 2020/21.
The figure reveals that passes depend on the formations of teams.
Figure \ref{fig:network-juventus} in Supplement \ref{sec:plots} shows that the midfield players and defenders of Juventus Turin (with 4-4-2 formation) dominate ball control.
By contrast,
strikers do not control the ball all too often,
but are key to scoring goals and hence winning matches.
Figure \ref{fig:network-inter} in Supplement \ref{sec:plots} reveals that the midfield players of Inter Milan (with 3-5-2 formation) likewise dominate ball control.
In addition,
the right wing of Inter Milan plays an important role in Inter Milan's 3-5-2 formation,
by passing the ball to the strikers and in so doing helping the team launch counterattacks straight out of the \textcolor{black}{backfield.}
Other descriptive summaries,
including detailed information on the formations and players of Juventus Turin, Inter Milan, and other soccer clubs in Serie A are presented in Supplement \ref{sec:descriptives}.

\s

\section{Stochastic modeling framework}
\label{sec:model}

We introduce a continuous-time stochastic process as a model of soccer matches starting at time $t_0 \coloneqq 0$ and stopping at time $T \in [90, +\infty)$.

Soccer matches involve two competing teams.
Each team consists of $11$ players and can substitute up to $5$ players during a match,
effective 2022/23.
Let $\mT_{1,t}$ be the set of players of one of the two teams and $\mT_{t,2}$ be the set of players of the opposing team at time $t \in [0,\, T)$.
The two sets $\mT_{1,t}$ and $\mT_{2,t}$ are disjoint,
in the sense that $\mT_{1,t}\, \cap\, \mT_{2,t} = \{\}$ for all $t \in [0,\, T)$.
The compositions of the two teams $\mT_{1,t}$ and $\mT_{2,t}$ can change during a match,
because players may be injured;
players may be substituted;
and the referee may remove players from the \textcolor{black}{field} due to violations of rules.
We consider changes in the compositions of $\mT_{1,t}$ and $\mT_{2,t}$ to be exogenous.

\subsection{Generic continuous-time stochastic process}
\label{basic}

We introduce a generic continuous-time stochastic process that captures salient features of soccer matches.\s

{\bf Scoring goals: rare events.}
We focus on who is control of the ball,
whether a change in ball control is a failure or a success,
and who secures control of the ball,
but we do not model the process of scoring goals.
While scoring goals is important for winning matches,
the event of scoring a goal is a rare event and useful predictors are hard to come by,
because scoring a goal requires a sequence of complex interactions among players of two competing teams.
We leave the construction of models for scoring goals to future research and focus here on ball control and interactions among players,
which are important for scoring goals and winning matches.\s

{\bf Ball control and interactions among players.}
We first describe a generic continuous-time stochastic process.
We then introduce a specification of the continuous-time stochastic process in Section \ref{spec} and discuss \textcolor{black}{basic} properties of the continuous-time stochastic process in Section \ref{markov}.\s

A generic continuous-time stochastic process of a soccer match starting at time $t_0 \coloneqq 0$ and stopping at time $T \in [90, +\infty)$ takes the following form:
\begin{enumerate}\item[1.] 
At time $t_0 \coloneqq 0$, 
the referee starts the match.
The player who secures control of the ball at time $t_0$ is chosen at random from the set $\mT_{1,t_0}\, \cup\; \mT_{2,t_0}$ and is denoted by $i_1$.
\item[2.] At time $t_m \coloneqq t_{m-1} + h_m$ ($m = 1, 2, \dots$),
the ball passes from player $i_m \in \mT_{1,t_m}\, \cup\; \mT_{2,t_m}$ to player $j_m \in \mT_{1,t_m}\, \cup\; \mT_{2,t_m} \setminus\, \{i_m\}$,
where $h_m \sim \mbox{Exponential}(\lambda_{i_m})$ and $i_m = j_{m-1}$ 
($m = 2, 3, \dots$).
The process of passing the ball from player $i_m$ to player $j_m$ is decomposed as follows:
\begin{enumerate}
\item[2.1] The change in ball control is either a failure (indicated by $S_{i_m} = 0$) in that player $i_m$ loses the ball to a player of the opposing team,
or is a success (indicated by $S_{i_m} = 1$) in that $i_m$ succeeds in passing the ball to a player of $i_m$'s own team.
\item[2.2] Conditional on $S_{i_m} \in \{0, 1\}$,
player $i_m$ cedes control of the ball to player $j_m \in \mT_{1,t_m} \cup \mT_{2,t_m} \setminus\, \{i_m\}$,
indicated by $i_m \to j_m$. 
\end{enumerate}
\item[3.] The referee stops the match at time $T \in [90, +\infty)$.
\end{enumerate}
We consider the decision of the referee to stop the match to be exogenous,
so that the stopping time $T \in [90, +\infty)$ of the match is non-random.
In practice,
soccer matches last 90 minutes,
but disruptions of matches due to injuries and substitutions of players may result in overtime.

\subsection{Specification of continuous-time stochastic process}
\label{spec}

We introduce a specification of the generic continuous-time stochastic process introduced in Section \ref{basic},
by specifying the distributions of the holding times $h_m$,
the success probabilities $\mbP(S_{i_m} = s_{i_m})$, 
and the pass probabilities $\mbP(i_m \to j_m \mid S_{i_m} = s_{i_m})$.
\textcolor{black}{Basic} properties of the resulting continuous-time stochastic process are discussed in Section \ref{markov}.
Throughout,
we denote by $\mathcal{I}_m$ the team of player $i_m$ in control of the ball at time $t_m$.
\hide{
that is,
$\mathcal{I}_m \coloneqq \mT_{1,t_m}$ if $i_m \in \mT_{1,t_m}$ and $\mathcal{I}_m \coloneqq \mT_{2,t_m}$ otherwise.
}
 
\paragraph{Holding time distributions}

A natural specification of the holding time distributions is 
\beno
h_m \mid \lambda_{i_m} &\ind& \mbox{Exponential}(\lambda_{i_m}).
\ee
To allow the rate $\lambda_{i_m} \in (0, +\infty)$ of player $i_m$'s holding time $h_m$ to depend on observed attributes of $i_m$ (e.g.,
the position of $i_m$ in the formation of $i_m$'s team and the location of $i_m$ on the field),
we assume that 
\beno
\lambda_{i_m}(\bm\omega)
&\coloneqq& \exp(\bm\omega^\top \bm{c}_{i_m}),
\ee
where $\bm\omega \in \mR^p$ is a vector of $p$ parameters and $\bm{c}_{i_m} \in \mR^p$ is a vector of $p$ observed attributes of player $i_m$.

\paragraph{Success probabilities}

The probability of a successful pass $\{S_{i_m} = 1\}$ by player $i_m$ can be specified by a logit model:
\beno
\logit(\mbP_{\bm\alpha,\bm\eta}(S_{i_m} = 1)) 
&\coloneqq& \bm\alpha^\top \bm{x}_{1,i_m} + \eta_{1,i_m},
\ee
where $\bm\alpha \in \mR^{d_1}$ is a vector of $d_1$ parameters and $\bm{x}_{1,i_m} \in \mR^{d_1}$ is a vector of $d_1$ observed attributes of player $i_m$.
The random effect $\eta_{1,i_m} \in \mR$ captures the effect of unobserved attributes of player $i_m$ on the success probability.

\paragraph{Pass probabilities}

The conditional probability of event $\{i_m \to j_m\}$ given $\{S_{i_m} = 0\}$ can be specified by a multinomial logit model:
\beno
\mbP_{\bm\beta,\bm\eta}(i_m \to j_m \mid S_{i_m} = 0)\s
\\
\coloneqq\;
\left\{\begin{array}{lll}
\dfrac{\exp(\bm\beta^\top \bm{x}_{2,i_m,j_m} + \eta_{2,j_m})}{\sum_{j\, \not\in\, \mathcal{I}_m} \exp(\bm\beta^\top \bm{x}_{2,i_m,j} + \eta_{2,j})} & \mbox{if } j_m \not\in \mathcal{I}_m
\\
0 & \mbox{if } j_m \in \mathcal{I}_m,
\end{array}\right\}
\ee
where $\bm\beta \in \mR^{d_2}$ is a vector of $d_2$ parameters and $\bm{x}_{2,i_m,j} \in \mR^{d_2}$ is a vector of $d_2$ observed attributes of players $i_m$ and $j$.
The random effect $\eta_{2,j} \in \mR$ captures the effect of unobserved attributes of player $j$ on the conditional probability of securing control of the ball.

{Along the same lines,
the conditional probability of event $\{i_m \to j_m\}$ given $\{S_{i_m} = 1\}$ can be specified by a multinomial logit model:
\beno
\mbP_{\bm\gamma,\bm\eta}(i_m \to j_m \mid S_{i_m} = 1)\s
\\
\coloneqq\;
\left\{\begin{array}{lll}
0 & \mbox{if } i_m = j_m \mbox{ or } j_m \not\in \mathcal{I}_m
\\
\dfrac{\exp(\bm\gamma^\top \bm{x}_{3,i_m,j_m} + \eta_{3,j_m})}{\sum_{j\, \in\, \mathcal{I}_m \setminus \{i_m\}} \exp(\bm\gamma^\top \bm{x}_{3,i_m,j} + \eta_{3,j})} & \mbox{if } i_m \neq j_m \mbox{ and } j_m \in \mathcal{I}_m,
\end{array}\right.
\ee
where $\bm\gamma \in \mR^{d_3}$ is a vector of $d_3$ parameters and $\bm{x}_{3,i_m,j} \in \mR^{d_3}$ is a vector of $d_3$ observed attributes of players $i_m$ and $j$,
e.g.,
whether player $i_m$ passed the ball to player $j$ in the past,
whether player $i_m$ received the ball from player $j$ in the past,
or the spatial distance between players $i_m$ and $j$ at the time of the pass (provided that the spatial positions of players are known).
The random effect $\eta_{3,j} \in \mR$ captures the effect of unobserved attributes of player $j$ on the conditional probability of securing control of the ball.
}

\paragraph{Random effects}

Let $\bm\eta_i \coloneqq (\eta_{1,i},\, \eta_{2,i},\, \eta_{3,i}) \in \mR^3$ and assume that 
\beno
\bm\eta_i \mid \bSigma &\iid& \mbox{MVN}_3(\bm{0}_3,\, \bSigma),
\ee
where $\bm{0}_3 \in \mR^3$ is the three-dimensional null vector and $\bSigma \in \mR^{3 \times 3}$ is a positive-definite variance-covariance matrix.

\s\s

\noindent
{\bf Alternative models}\;\;
{It is worth noting that there are other possible approaches to constructing stochastic models of soccer matches.
For example,
each the two following approaches to constructing models can help shed light on salient aspects of soccer matches:
\begin{itemize}
\item[(a)] Assuming player $i_m$ is in control of the ball at time $t_m$, 
first determine whether $i_m$ succeeds in passing the ball to a teamplayer.
Then determine which teamplayer $j_m$ receives the ball provided that the pass is a success,
otherwise determine which player $j_m$ of the opposing team secures control of the ball provided that the pass is a failure (the approach pursued here).
\item[(b)] Assuming player $i_m$ is in control of the ball at time $t_m$,
suppose that $i_m$ first selects a teamplayer $k_m$ and intends to pass the ball to $k_m$.
Then determine whether the intended pass $i_m \to k_m$ succeeds.
If the intended pass $i_m \to k_m$ succeeds,
set $k_m = j_m$,
otherwise select the player $j_m$ who secures control of the ball from the opposing team (an approach suggested by an anonymous referee).
\end{itemize}
While both approaches can be useful,
there are two good reasons for choosing approach (a),
that is,
the approach pursued here.

First,
soccer matches revolve around the ball,
so soccer teams wish to retain control of the ball.
Thus,
the player in control of the ball is first and foremost responsible for passing the ball to a teamplayer---unless the player has the rare opportunity to score a goal.
By construction,
approach (a) respects the importance of retaining control of the ball.

Second,
approach (a) has one advantage over approach (b):
If a pass is a failure, 
we do not observe the intended receiver $k_m$.
Worse,
even when a pass is a success,
we may not observe the intended receiver $k_m$:
e.g.,
$i_m$ may intend to pass the ball to teamplayer $k_m$,
but the ball ends up in possession of some other teamplayer $j_m \neq k_m$ by accident.
In fact,
instead of observing the intended receiver $k_m$,
we observe the actual receiver $j_m$,
who may or may not be identical to the intended receiver $k_m$.
In other words,
the data fall short,
in that we do not observe the {\it intended passes $i_m \to k_m$,}
but we observe the {\it actual passes $i_m \to j_m$,} 
regardless of whether the passes are failures or successes.
As a result,
approach (b) would require augmenting the observed passes $i_m \to j_m$ by the unobserved, intended passes $i_m \to k_m$.
While it is possible to augment the observed passes $i_m \to j_m$ by the unobserved, intended passes $i_m \to k_m$ using data-augmentation methods,
such methods come at additional computational costs compared with approach (a).
In addition,
there may be statistical costs:
It is not clear how much information the data contain about the unobserved, intended passes $i_m \to k_m$.

}

\subsection{Properties of continuous-time stochastic process}
\label{markov}

We discuss basic properties of the continuous-time stochastic process specified in Sections \ref{basic} and \ref{spec}.
Throughout Section \ref{markov},
we suppress the notational dependence of all quantities on the parameters $\bm\alpha,\, \bm\beta,\, \bm\gamma,\, \bm\omega,\, \bm\Sigma$ and the random effects $\bm\eta_1, \bm\eta_2, \dots${We assume that the continuous-time stochastic process satisfies two assumptions:
\bi
\item[A.1] In a short time interval $[t_1,\, t_2]$,
the compositions of teams $\mT_{1,t}$ and $\mT_{2,t}$ are constant, 
in the sense that $\mT_{1,t} \equiv \mT_1$ and $\mT_{2,t} \equiv \mT_2$ for all $t \in [t_1,\, t_2)$, 
and the 22 players of the two teams are labeled $1, \dots, 22$.
\item[A.2] In a short time interval $[t_1,\, t_2]$,
the attributes of players and teams,
the rates $\lambda_i$,
the success probabilities $\mbP(S_i = s_i)$,
and the pass probabilities $\mbP(i \to j \mid S_i = s_i)$ are time-invariant.
\ei

Assumption A.1 states that the compositions of the teams do not change in a short time interval,
that is,
the two teams do not substitute players.
Assumption A.2 ensures that the continuous-time stochastic process is time-homogeneous in a short time interval.
The assumption that the continuous-time stochastic process is time-homogeneous in a short time interval is not unreasonable,
because soccer teams consist of humans,
and humans are incapable of instantaneous changes.
We hasten to point out that the stochastic modeling framework is not restricted to time-homogeneous stochastic processes: 
It does allow the attributes of players and teams,
the rates $\lambda_i$,
the success probabilities $\mbP(S_i = s_i)$,
and the pass probabilities $\mbP(i \to j \mid S_i = s_i)$ to change over time.
The purpose of the following proposition is to shed light on the behavior of the continuous-time stochastic process in a short time interval,
during which the stochastic process can be approximated by a time-homogeneous stochastic process.
}

\begin{proposition}
Consider the continuous-time stochastic process described in Sections \ref{basic} and \ref{spec} satisfying Assumptions A.1 and A.2.
Then the stochastic process is a right-continuous and time-homogeneous Markov process $\{Y(t),\, t \in [t_1,\, t_2)\}$ with finite state space $\mathscr{Y} \coloneqq \{1, \dots, 22\}$ during a short time interval $[t_1, t_2)$,
where the state $Y(t) \in \mathscr{Y}$ of the Markov process at time $t$ indicates which player is in control of the ball at time $t$.
The elements $q_{i,j}$ of the generator matrix $\bm{Q} \in \mR^{|\mathscr{Y}| \times |\mathscr{Y}|}$ of the Markov process are
\vspace{-1cm}
\beno
q_{i,j}
&\coloneqq&
\left\{\begin{array}{lll}
\lambda_{i}\;\, \mbP(S_i = 0)\;\, \mbP(i \to j \mid S_i = 0) 
& \mbox{if } i \neq j \mbox{ and } j \not\in \mathcal{I}_i
\\
\lambda_{i}\;\, \mbP(S_i = 1)\;\, \mbP(i \to j \mid S_i = 1)
& \mbox{if } i \neq j \mbox{ and } j \in \mathcal{I}_i
\\
-\lambda_{i} & \mbox{if } i = j,
\end{array}\right\}
\ee
where $\mathcal{I}_i$ denotes the team of player $i \in \mathscr{Y}$.
Consider any $t \in [t_1,\, t_2)$ and any $h \in (0,\, t_2 - t)$.
Then,
for all $(i, j) \in \mathscr{Y}^2$,
conditional on $\{Y(t) = i\}$,
the event $\{Y(t + h) = j\}$ is independent of $\{Y(s),\, s \leq t\}$ and,
as $h \downarrow 0$,
the conditional probability of event $\{Y(t + h) = j\}$ given $\{Y(t) = i\}$ is
\beno
\mbP(Y(t + h) = j \mid Y(t) = i)
&=& \delta_{i,j} + q_{i,j}\, h + o(h),
\ee 
where $\delta_{i,j} \coloneqq 1$ if $i = j$ and $\delta_{i,j} \coloneqq 0$ otherwise.
\end{proposition}

{The proposition is a straightforward consequence of the construction of the continuous-time stochastic process and Theorem 2.8.2 of\; \citet[p.\ 94]{Nj97}.
The proposition shows that the continuous-time stochastic process focuses on ball control and who passes the ball to whom,
by specifying the rates $q_{i,j}$ of passing the ball between pairs of players $(i, j) \in \mathcal{Y}^2$.
It implies that the conditional probability of the event that player $j$ is in possession of the ball at time $t + h$, 
given that player $i \neq j$ is in possession of the ball at time $t$, 
is approximately $q_{i,j}\, h$ in a short time interval of length $h$.
}

\section{Bayesian learning}
\label{sec:inference}

We pursue a Bayesian approach to learning the stochastic modeling framework introduced in Section \ref{sec:model} from high-resolution network data.

A Bayesian approach is well-suited to online learning, 
that is,
updating the knowledge about the parameters $\bm\alpha,\, \bm\beta,\, \bm\gamma,\, \bm\omega,\, \bm\Sigma$ and the random effects $\bm\eta_1, \bm\eta_2, \dots$ as soon as additional data points roll in.
To demonstrate,
consider two teams and let $\bm{x}_1 \coloneqq (h_{1,m},\, i_{1,m},\, j_{1,m})_{m=1}^{M_1}$ be the outcome of the first match of the two teams (with $M_1 \geq 1$ passes) and $\bm{x}_2 \coloneqq (h_{2,m},\, i_{2,m},\, j_{2,m})_{m=1}^{M_2}$ be the outcome of the second match of the two teams (with $M_2 \geq 1$ passes).
To ease the presentation,
assume that the compositions of the two teams do not change during the first and second match,
the 22 players of the two teams are labeled $1, \dots, 22$,
and the random effects are denoted by $\bm\eta \coloneqq (\bm\eta_1, \dots, \bm\eta_{22})$.
In addition,
assume that the outcomes of the first and second match $\bm{x}_1$ and $\bm{x}_2$ satisfy
\beno
\pi(\bm{x}_1,\, \bm{x}_2 \mid \bm\alpha,\, \bm\beta,\, \bm\gamma,\, \bm\omega,\, \bm\eta)
&=& \pi(\bm{x}_1 \mid \bm\alpha,\, \bm\beta,\,  \bm\gamma,\, \bm\omega,\, \bm\eta)\\
&\times& \pi(\bm{x}_2 \mid \bm\alpha,\, \bm\beta,\, \bm\gamma,\, \bm\omega,\, \bm\eta,\, \bm{x}_1),
\ee
where $\pi$ denotes a generic probability density function.
The conditional probability density function $\pi(\bm{x}_1 \mid \bm\alpha,\, \bm\beta,\,  \bm\gamma,\, \bm\omega,\, \bm\eta)$ is of the form
\beno
\pi(\bm{x}_1 \mid \bm\alpha,\, \bm\beta,\,  \bm\gamma,\, \bm\omega,\, \bm\eta)
\;=\; \dprod_{m=1}^{M_1}\, \Big[\lambda_{i_{1,m}}(\bm\omega)\, \exp(-\lambda_{i_{1,m}}(\bm\omega)\, h_{1,m})\s
\\
\hspace{4.75cm}\times\;\; \mbP_{\bm\alpha,\bm\eta}(S_{i_{1,m}} = s_{i_{1,m}})\s
\\
\hspace{4.75cm}\times\;\; \mbP_{\bm\beta,\bm\eta}(i_{1,m} \to j_{1,m} \mid S_{i_{1,m}} = 0)^{\one(S_{i_{1,m}} =\; 0)}\s
\\
\hspace{4.75cm}\times\;\; \mbP_{\bm\gamma,\bm\eta}(i_{1,m} \to j_{1,m} \mid S_{i_{1,m}} = 1)^{\one(S_{i_{1,m}} =\; 1)}\Big]\s\s
\\
\hspace{4.75cm}\times\;\; \exp\left(-\lambda_{i_{1,M_1+1}}(\bm\omega)\, \left(T_1 - \dsum_{k=1}^{M_1} h_{1,k}\right)\right),
\ee
assuming that the start time $t_0 \coloneqq 0$ and the stopping time $T_1 \in [90, +\infty)$ of the match are determined by the referee and are both non-random.
The function $\one(.)$ is an indicator function,
which is $1$ if its argument is true and is $0$ otherwise.
The conditional probability density function $\pi(\bm{x}_2 \mid \bm\alpha,\, \bm\beta,\, \bm\gamma,\, \bm\omega,\, \bm\eta,\, \bm{x}_1)$ is of the same form as $\pi(\bm{x}_1 \mid \bm\alpha,\, \bm\beta,\,  \bm\gamma,\, \bm\omega,\, \bm\eta)$,
but is based on $M_2$ passes rather than $M_1$ passes and can depend on the outcome of the first match $\bm{x}_1$.

The posterior of $\bm\alpha,\, \bm\beta,\, \bm\gamma,\, \bm\omega,\, \bm\Sigma,\, \bm\eta$ based on the outcome of the first match $\bm{x}_1$ is proportional to
\[
\begin{array}{llllllllll}
\pi(\bm\alpha,\, \bm\beta,\, \bm\gamma,\, \bm\omega,\, \bm\Sigma,\, \bm\eta \mid \bm{x}_1)
&\propto& \pi(\bm{x}_1 \mid \bm\alpha,\, \bm\beta,\, \bm\gamma,\, \bm\omega,\, \bm\eta)\\
&\times& \pi(\bm\eta \mid \bm\Sigma)\; \pi(\bm\alpha,\, \bm\beta,\, \bm\gamma,\, \bm\omega,\, \bm\Sigma),
\end{array}
\]
where $\pi(\bm\alpha,\, \bm\beta,\, \bm\gamma,\, \bm\omega,\, \bm\Sigma)$ is the prior of $\bm\alpha,\, \bm\beta,\, \bm\gamma,\, \bm\omega,\, \bm\Sigma$.
The prior of $\bm\alpha,\, \bm\beta,\, \bm\gamma,\, \bm\omega,\, \bm\Sigma$ is described in Section \ref{sec:computing}.

As soon as the outcome of the second match $\bm{x}_2$ is observed,
the knowledge about $\bm\alpha,\, \bm\beta,\, \bm\gamma,\, \bm\omega,\, \bm\Sigma,\, \bm\eta$ in light of $\bm{x}_2$ can be updated as follows:
\beno
\label{posterior.updating}
\begin{array}{llllllllll}
&& \pi(\bm\alpha,\, \bm\beta,\, \bm\gamma,\, \bm\omega,\, \bm\Sigma,\, \bm\eta \mid \bm{x}_1,\, \bm{x}_2)
\\
&\propto& \pi(\bm{x}_1,\, \bm{x}_2 \mid \bm\alpha,\, \bm\beta,\, \bm\gamma,\, \bm\omega,\, \bm\eta)\; \pi(\bm\eta \mid \bm\Sigma)\; \pi(\bm\alpha,\, \bm\beta,\, \bm\gamma,\, \bm\omega,\, \bm\Sigma)
\\
\hide{
&\propto& \pi(\bm{x}_2 \mid \bm\alpha,\, \bm\beta,\, \bm\gamma,\, \bm\Sigma,\, \bm\omega,\, \bm{x}_1)\;\pi(\bm{x}_1 \mid \bm\alpha,\, \bm\beta,\, \bm\gamma,\, \bm\Sigma,\, \bm\omega)\, \times\, \pi(\bm\alpha,\, \bm\beta,\, \bm\gamma,\, \bm\Sigma,\, \bm\omega)\s\s 
\\
}
&\propto& \pi(\bm{x}_2 \mid \bm\alpha,\, \bm\beta,\, \bm\gamma,\, \bm\omega,\, \bm\eta,\, \bm{x}_1)\; {\pi(\bm\alpha,\, \bm\beta,\, \bm\gamma,\, \bm\omega,\, \bm\Sigma,\, \bm\eta \mid \bm{x}_1)}.
\end{array}
\ee
In other words,
as soon as the outcome of the second match $\bm{x}_2$ is observed,
we can update the knowledge about $\bm\alpha,\, \bm\beta,\, \bm\gamma,\, \bm\omega,\, \bm\Sigma,\, \bm\eta$ in light of $\bm{x}_2$ via $\pi(\bm{x}_2 \mid \bm\alpha,\, \bm\beta,\, \bm\gamma,\, \bm\omega,\, \bm\eta,\, \bm{x}_1)$,
with the knowledge about $\bm\alpha,\, \bm\beta,\, \bm\gamma,\, \bm\omega,\, \bm\Sigma,\, \bm\eta$ prior to the second match $\bm{x}_2$ being quantified by $\pi(\bm\alpha,\, \bm\beta,\, \bm\gamma,\, \bm\omega,\, \bm\Sigma,\, \bm\eta \mid \bm{x}_1)$,
the posterior based on the outcome of the first match $\bm{x}_1$.
As a result,
a Bayesian approach is a natural approach to updating knowledge about the stochastic modeling framework as additional data points roll in.
More than two teams with can be handled,
and multiple matches in parallel.
{\section{Bayesian computing}
\label{sec:computing}
}

{
The posterior $\pi(\bm\alpha,\, \bm\beta,\, \bm\gamma,\, \bm\omega,\, \bm\Sigma,\, \bm\eta \mid \bm{x})$ of the parameters $\bm\alpha,\, \bm\beta,\, \bm\gamma,\, \bm\omega,\, \bm\Sigma$ and the random effects $\bm\eta$ based on the outcome of a match $\bx$ is not available in closed form.
We approximate the posterior by using Markov chain Monte Carlo methods,
by sampling from the full conditional distributions of the parameters and the random effects:
\beno
\pi(\bm\alpha \mid \bx) 
&\propto& \mathscr{L}(\bm\alpha, \bm\eta;\, \bx)\, \pi(\bm\alpha)\s
\\
\pi(\bm\beta \mid \bx) 
&\propto& \mathscr{L}(\bm\beta, \bm\eta;\, \bx)\, \pi(\bm\beta)\s
\\
\pi(\bm\gamma \mid \bx) 
&\propto& \mathscr{L}(\bm\gamma, \bm\eta;\, \bx)\, \pi(\bm\gamma)\s
\\
\pi(\bm\omega \mid \bx) 
&\propto& \mathscr{L}(\bm\omega;\, \bx)\, \pi(\bm\omega)\s
\\ 
\pi(\bm\eta \mid \bx)
&\propto& \mathscr{L}(\balpha, \bm\eta;\, \bx)\, \mathscr{L}(\bbeta, \bm\eta;\, \bx)\, \mathscr{L}(\bm\gamma, \bm\eta;\, \bx)\, \mathscr{L}(\bSigma;\, \bm\eta)\s
\\ 
\pi(\bm\Sigma \mid \bm\eta) 
&\propto& \mathscr{L}(\bSigma;\, \bm\eta)\, \pi(\bm\Sigma), 
\label{eq: cond distributions}
\ee
where
\beno
\mathscr{L}(\balpha, \bm\eta;\, \bx)
&\propto& \dprod_{m=1}^{M}\, \mbP_{\bm\alpha,\bm\eta}(S_{i_{m}} = s_{i_{m}})\s
\\
\mathscr{L}(\bbeta, \bm\eta;\, \bx)
&\propto& \dprod_{m=1}^{M}\,  \mbP_{\bm\beta,\bm\eta}(i_{m} \to j_{m} \mid S_{i_{m}} = 0)^{\one(S_{i_{m}} =\; 0)}\s
\\
\mathscr{L}(\bm\gamma, \bm\eta;\, \bx)
&\propto& \dprod_{m=1}^{M}\,  \mbP_{\bm\gamma,\bm\eta}(i_{m} \to j_{m} \mid S_{i_{m}} = 1)^{\one(S_{i_{m}} =\; 1)}\s
\\
\mathscr{L}(\bm\omega;\, \bx)
&\propto& \dprod_{m=1}^{M}\, \left[\lambda_{i_{m}}(\bm\omega)\, \exp(-\lambda_{i_{m}}(\bm\omega)\, h_{m})\right]\,  \exp\left(-\lambda_{i_{M+1}}(\bm\omega)\, \left(T - \dsum_{k=1}^{M} h_{k}\right)\right)\s
\\
\mathscr{L}(\bSigma;\, \bm\eta) 
&\propto& \dprod_{i=1}^n\, \mbox{det}({\bm\Sigma^{-1}})^{1/2}\, \exp\left(-\dfrac{1}{2}\, \bm\eta_i^\top\, \bm\Sigma^{-1}\, \bm\eta_i\right),
\ee
assuming that $\bx$ is the outcome of a single soccer match with $M \geq 1$ passes starting at time $t_0 = 0$ and stopping at time $T \in [90, +\infty)$;
note that both the start time $t_0$ and the stopping time $T$ are non-random.
We assume that the prior factorizes according to
\beno
\pi(\bm\alpha,\, \bm\beta,\, \bm\gamma,\, \bm\omega,\, \bm\Sigma)
&=& \pi(\bm\alpha)\, \pi(\bm\beta)\, \pi(\bm\gamma)\, \pi(\bm\omega)\, \pi(\bm\Sigma),
\ee
with marginal priors of the form
\beno
\alpha_k &\iid& N(0, 10^2), & k = 1, \dots, d_1, && \beta_k &\iid& N(0, 10^2), & k = 1, \dots, d_2
\\
\gamma_k &\iid& N(0, 10^2), & k = 1, \dots, d_3, && \omega_k &\iid& N(0, 10^2), & k = 1, \dots, p,
\ee
where $N(0, 10^2)$ is a Gaussian with mean $0$ and variance $10^2 = 100$.
To specify the prior of the variance-covariance matrix $\bm\Sigma$ of the random effects,
we decompose $\bm\Sigma$ according to
\beno
\bSigma &\coloneqq& 
\left(
\begin{array}{llll}
        \sigma_{\eta_1}& 0 & 0\\
        0 & \sigma_{\eta_2} & 0\\
0 & 0 & \sigma_{\eta_3}
\end{array}
\right)\,
\bm{\Lambda} 
\left(
\begin{array}{llll}
        \sigma_{\eta_1}& 0 & 0\\
        0 & \sigma_{\eta_2} & 0\\
0 & 0 & \sigma_{\eta_3}
\end{array}
\right),
\ee
where $\bm\Lambda \in [-1,+1]^{3 \times 3}$ is a correlation matrix.
We then assume that $\bm\Lambda \sim \mbox{LKJcorr}(2)$ has a Lewandowski-Kurowicka-Joe (LKJ) distribution with parameter $2$ and $\sigma_{\eta_k} \iid \mbox{Exponential}(1)$ ($k = 1, 2, 3$).

To sample from the full conditionals,
we use Markov chain Monte Carlo methods implemented in {\tt R} package {\tt rstan} \citep{rstan}.
Since the stochastic modeling framework leverages exponential-family distributions as building blocks (e.g., Bernoulli, Multinomial, Exponential, and multivariate Gaussians),
we do not have more numerical issues than other exponential-family models,
such as generalized linear models,
Gaussian and non-Gaussian graphical models \citep{efron22}.

}

\s\s

\section{Application}
\label{sec:app}

We use the stochastic modeling framework introduced in Section \ref{sec:model} to analyze the data described in Section \ref{sec:data}.
We focus on the matches of four soccer teams during the 2020/21 season of Serie A,
the premier league of the Italian football league system:
\bi
\item Juventus Turin (Juventus F.C.;
15,832 observations);
\item Inter Milan (Internazionale Milano;
13,564 observations);
\item Crotone (Crotone S.r.l.;
8,125 observations);
\item Fiorentina (ACF Fiorentina;
8,107 observations).
\ei
Juventus Turin and Inter Milan belong to the most storied Italian soccer clubs,
while Crotone and Fiorentina were mid- and low-level teams during the 2020/21 season,
respectively.
The numbers of observations mentioned above refer to the total numbers of passes during the 2020/21 season,
aggregated over all matches played by the selected teams with the dominant formation.
The selected teams have in common that all of them were proficient users of the 4-4-2  formation (Juventus Turin) or the 3-5-2 formation (Inter Milan, Crotone, Fiorentina).

We use the following specification of the stochastic modeling framework:
\bi
\item {\bf Module 1 (M1):} 
The Exponential model of the holding times $h_m$ uses the following covariates:
position-specific indicators of who is in control of the ball and indicators of whether the player's team is on track to winning or losing the match (i.e., the player's team has scored at least one more goal or one less goal than its opponent, 
respectively).
\item {\bf Module 2 (M2):} 
The logit model of the probability of a successful pass $\{S_{i_m} = 1\}$ uses the following covariates,
in addition to an intercept:
the length of the pass in terms of two-dimensional Euclidean distance;
an indicator of whether player $i_m$ initiates the pass in the opposing team's half of the field;
an indicator of whether the ball ends up in the opposing team's third of the field;
an indicator of whether the pass is a forward pass;
an indicator of whether the pass is an air pass; 
indicators of whether the player's team is on track to winning or losing the match (i.e., whether the player's team has scored at least one more goal or one less goal than its opponent, 
respectively);
and a position-specific random effect.
\item {\bf Module 3 (M3):} 
The multinomial logit model of the conditional probability of event $\{i_m \to j_m\}$ given $\{S_{i_m} = 1\}$ uses the following predictors:
the graph distance between players $i_m$ and $j_m$---defined as the length of the shortest path between $i_m$ and $j_m$---based on the nearest-neighbor graph in Figure \ref{nngraph2} in Supplement \ref{sec:plots};
the number of times $j_m$ received the ball prior to the $m$-th pass;
and a position-specific random effect.
\ei
{It would be interesting to include more features into the multinomial logit model of the conditional probability of event $\{i_m \to j_m\}$ given $\{S_{i_m} = 1\}$, 
e.g.,
the spatial positions of players and additional network features.
That said,
we do not have data on the spatial distances between players and additional network features.
Note that these limitations are limitations of the data,
not the model:
The model can incorporate spatial distances between players as well as additional network features.
}In addition,
note that we focus here on all matches involving the four mentioned teams with the dominant formation,
but we do not use the data of the opposing teams.
As a consequence,
we do not specify the conditional probabilities of events $\{i_m \to j_m\}$ given $\{S_{i_m} = 0\}$.
Last,
but not least,
note that we use position-specific rather than player-specific random effects, 
because the data do not include complete information about which position is filled by which player.

{Posterior sensitivity checks and posterior predictive checks can be found in Sections \ref{sec:sensitivity} and \ref{sec:predictions},
respectively:
The posterior sensitivity checks suggest that the posterior is not too sensitive to the choice of prior,
while the posterior predictive checks indicate that model-based predictions match the observed data.
Tables~\ref{tab:appRed} and~\ref{tab:appRed_juventus} in Supplement \ref{sec:summaries} present posterior summaries of the model parameters, 
based on the 2020/21 matches of Fiorentina, Crotone, and Inter Milan (with 3-5-2 formation) and Juventus Turin (with 4-4-2 formation).
Among other things,
these results suggest that all teams reduce the pace of passing the ball when being on track to winning a match. 
By contrast,
when on track to losing a match,
Juventus Turin and Inter Milan reduce the pace,
whereas Fiorentina and Crotone do not.
These results are surprising.
As a case in point,
consider Inter Milan:
When Inter Milan is on track to losing a match,
one would expect that Inter Milan either maintains or increases its pace to change the outcome of the match in its favor.
But the results suggest otherwise:
In the face of defeat,
Inter Milan appears to reduce its pace---regardless of whether the opposing team is weak or strong.
Juventus Turin appears to do the same.
By contrast,
Fiorentina and Crotone maintain their pace when on track to losing a match.
The difference suggests that the strategies of Juventus Turin and Inter Milan for dealing with adverse situations differ from the strategies of Fiorentina and Crotone.
There is an additional observation suggesting that the strategies of Juventus Turin and Inter Milan differ from the strategies of Fiorentina and Crotone:
Starting a pass in the opponent's half of the field does not increase the probability of a successful pass among Fiorentina and Crotone players,
but it does increase the probability of a successful pass among Juventus Turin and Inter Milan players.
The increase in the probability of a successful pass in the opponent's half of the field may be due to the offensive strength of Juventus Turin and Inter Milan.
Taken together,
these results suggest that the strategies of strong teams in the face of defeat differ from those of less strong teams: 
Strong teams may have the luxury to decrease their pace and leverage their offensive strength, 
while less strong teams may not be able to do so.
}

Among the position-specific effects,
it is worth noting that the length of time the goal keeper controls the ball tends to be lower than the length of time other positions control the ball.
This observation makes sense,
because the goal keeper has an incentive to remove the ball from the penalty area as soon as possible,
so that the opposing team cannot gain control of the ball in the penalty area and score an easy goal.
{\subsection{Posterior sensitivity checks}\label{sec:sensitivity}
}

{To assess the sensitivity of the posterior to the choice of prior,
we consider the following three priors:
\bi
\item Prior 1:
\beno
\alpha_k &\iid& N(0, 5^2), & k = 1, \dots, d_1, && \beta_k &\iid& N(0, 5^2), & k = 1, \dots, d_2
\\
\gamma_k &\iid& N(0, 5^2), & k = 1, \dots, d_3, && \omega_k &\iid& N(0, 5^2), & k = 1, \dots, p;
\ee
\item Prior 2,
used in Section \ref{sec:app}:
\beno
\alpha_k &\iid& N(0, 10^2), & k = 1, \dots, d_1, && \beta_k &\iid& N(0, 10^2), & k = 1, \dots, d_2
\\
\gamma_k &\iid& N(0, 10^2), & k = 1, \dots, d_3, && \omega_k &\iid& N(0, 10^2), & k = 1, \dots, p;
\ee
\item Prior 3:
\beno
\alpha_k &\iid& N(0, 15^2), & k = 1, \dots, d_1, && \beta_k &\iid& N(0, 15^2), & k = 1, \dots, d_2
\\
\gamma_k &\iid& N(0, 15^2), & k = 1, \dots, d_3, && \omega_k &\iid& N(0, 15^2), & k = 1, \dots, p;
\ee
\ei
where $N(0, 5^2)$, $N(0, 10^2)$, and $N(0, 15^2)$ are Gaussians with mean $0$ and variances $5^2 = 25$, $10^2 = 100$, and $15^2 = 225$,
respectively.
The random effects prior is described in Section \ref{sec:computing} and is the same under all three priors.
The posteriors under these three priors are similar,
as can be seen by comparing the tables in Supplements \ref{sec:summaries} and \ref{app:Sensitvity}.
}
{\subsection{Posterior predictive checks}\label{sec:predictions}
}

{Using the posterior draws generated in Section \ref{sec:app},
we compare model-based predictions of the waiting times between passes and the proportions of successful passes to the observed waiting times and the observed proportions of successful passes by Inter Milan, Crotone, and Fiorentina during the 2020/21 season.
The model-based predictions (i.e., posterior predictions) are shown in Figure \ref{fig:postpredcheck} in Supplement \ref{sec:posterior.predictive.checks} and match the observed data.}

{\section{Simulation results}
\label{sec:simulations}
}

{We simulate data from the stochastic modeling framework specified in Section \ref{sec:app}.
We choose the data-generating parameters of the model so that the simulated data mimic the Inter Milan data in Section \ref{sec:app}.
We simulate 100 short soccer seasons, 
each with 1,000 passes.
To estimate the model from the 100 simulated soccer seasons,
we leverage the Bayesian approach described in Section \ref{sec:computing}, 
using the prior described in Section \ref{sec:computing}.
We present in Figure \ref{ap:SimResults} in Supplement \ref{sec:simulation.results} aggregated simulations results based on all 100 simulated soccer seasons.
In addition,
we present the data-generating parameters along with posterior summaries of the parameters based on one of the 100 simulated soccer seasons in Table \ref{tab:appSim} in Supplement \ref{sec:simulation.results}.
The figure and table demonstrate that the posterior means of the parameters cluster around the data-generating parameters.
}

\s\s

\section{Discussion}
\label{disc}

\hide{

We have introduced a continuous-time stochastic modeling framework for studying interactions among players in soccer and other team-based sports.
While the proposed stochastic modeling framework is tailored to soccer matches,
many components of the framework can be adapted to other team-based sports (e.g., basketball),
though some of the components will have to be tailored to the specific application at hand.

}

We view the proposed stochastic modeling framework as a first step to modeling soccer matches and other team-based sports as space- and time-indexed network processes and hope that it will stimulate future research.
To stimulate future research,
we conclude with a short discussion of open questions and directions for future research.

\hide{

\subsection{Predicting goals and match outcomes}

The holy grail of sport statistics is to predict match outcomes.
In soccer,
the greatest obstacle to predicting goals and hence match outcomes is the fact that the event of scoring a goal is a rare event and useful predictors are hard to come by,
because scoring a goal requires a sequence of complex interactions among players of two competing teams.
As a first step,
we have therefore focused on ball control and interactions among players,
which are important for scoring goals and winning matches.
That said,
we hope that advances in data collection,
statistical modeling,
and statistical computing help understand and predict such rare events \citep{brechot2020dealing}.
\hide{
An additional challenge is the fact that even in an ideal world in which prefect predictions for such rare events were possible,
the predictions could be invalidated because teams react to predictions:
e.g.,
if team A is predicted to score a goal during the last 10 minutes of a match provided that the current compositions and formations of teams A and B do not change,
then team B has incentive to change its current composition and formation by, 
e.g.,
substituting fresh players for tired players and changing its formation to strengthen its defense.
}
 
}

\subsection{Model specification}

The deluge of high-resolution network data generated by soccer and other team-based sports implies that there are many possible features that may be relevant for predicting ball control,
goals, 
and match outcomes.
The specific features used in Section \ref{sec:app} make sense as a starting point,
but sound model selection procedures and more data are needed to shed light on which features are useful for predicting ball control,
goals,
and match outcomes. 

{In addition,
the proposed stochastic modeling framework includes player-specific random effects $\bm\eta_i \in \mR^3$,
which are correlated within players $i$ but are shared across soccer matches.
Since the proposed stochastic modeling framework is already fairly complex,
we stick to the player-specific random effects.
More advanced latent process models---e.g., 
multilevel models with position- and team-specific random effects and other more complex latent process models---constitute an interesting direction for future research.}
{\subsection{Data-related challenges}While there is a deluge of high-resolution network data,
countless data-related challenges remain.
For example,
while the stochastic modeling framework can incorporate the spatial distances between soccer players,
the existing data is limited to ball-tracking data and does not include player-tracking data.
As a result,
we know which player is in control of the ball,
but we do not know where the players are located on the field.
In the application,
we therefore used the graph distance based on the formations of soccer teams as an approximation of the spatial distances between players.
Note that these limitations are limitations of the data, 
not the model: 
The model can leverage both ball-tracking and player-tracking data.
Despite these limitations of existing data,
we believe that it is important to develop a stochastic modeling framework that can handle data that could and should be collected in the future.

\hide{
Other data-related challenges concern online learning.
While a Bayesian approach is well-suited to online learning,
there are at least two data-related challenges to online learning.
First,
current technology does not supply data without delay and without human intervention (i.e., without data cleaning by humans) that could be used by a Bayesian algorithm to update knowledge and make predictions.
Second,
Bayesian computing may not be fast enough to update knowledge about the stochastic process and make model-based predictions without delay.
}
}

\hide{

{We acknowledge support by the U.S.\ National Science Foundation in the form of NSF awards SES-2243058 and DMS-2412923 (GH) and 
the U.S.\ Department of Defense award ARO W911NF-21-1-0335 (MS).
We are indebted to two anonymous referees for numerous constructive suggestions that greatly improved the manuscript.}

\s

}

\bibliographystyle{chicago}

\bibliography{main}

\begin{thebibliography}{}

\bibitem[\protect\citeauthoryear{Albert, Glickman, Swartz, and Koning}{Albert
  et~al.}{2017}]{albert2017handbook}
Albert, J., M.~E. Glickman, T.~B. Swartz, and R.~H. Koning (2017).
\newblock {\em Handbook of Statistical Methods and Analyses in Sports}.
\newblock Boca Raton, FL: Chapman \& Hall/CRC.

\bibitem[\protect\citeauthoryear{Baio and Blangiardo}{Baio and
  Blangiardo}{2010}]{baio2010bayesian}
Baio, G. and M.~Blangiardo (2010).
\newblock Bayesian hierarchical model for the prediction of football results.
\newblock {\em Journal of Applied Statistics\/}~{\em 37\/}(2), 253--264.

\bibitem[\protect\citeauthoryear{Beal}{Beal}{2003}]{Be03}
Beal, M.~J. (2003).
\newblock {\em Variational Algorithms for Approximate Bayesian Inference}.
\newblock Ph.\ D. thesis, Gatsby Computational Neuroscience Unit, University
  College London.

\bibitem[\protect\citeauthoryear{Blei, Kucukelbir, and McAuliffe}{Blei
  et~al.}{2017}]{BlKuMA17}
Blei, D.~M., A.~Kucukelbir, and J.~D. McAuliffe (2017).
\newblock Variational inference: {A} review for statisticians.
\newblock {\em Journal of the American Statistical Association (Review)\/}~{\em
  112\/}(518), 859--877.

\bibitem[\protect\citeauthoryear{Brechot and Flepp}{Brechot and
  Flepp}{2020}]{brechot2020dealing}
Brechot, M. and R.~Flepp (2020).
\newblock Dealing with randomness in match outcomes: {H}ow to rethink
  performance evaluation in {E}uropean club football using expected goals.
\newblock {\em Journal of Sports Economics\/}~{\em 21\/}(4), 335--362.

\bibitem[\protect\citeauthoryear{Butts}{Butts}{2008}]{Butts2008}
Butts, C.~T. (2008).
\newblock A relational event framework for social action.
\newblock {\em Sociological Methodolgy\/}~{\em 38}, 155--200.

\bibitem[\protect\citeauthoryear{Cattelan, Varin, and Firth}{Cattelan
  et~al.}{2013}]{cattelan2013dynamic}
Cattelan, M., C.~Varin, and D.~Firth (2013).
\newblock Dynamic {B}radley--{T}erry modelling of sports tournaments.
\newblock {\em Journal of the Royal Statistical Society: Series C (Applied
  Statistics)\/}~{\em 62\/}(1), 135--150.

\bibitem[\protect\citeauthoryear{Cervone, D’Amour, Bornn, and
  Goldsberry}{Cervone et~al.}{2014}]{cervone2014pointwise}
Cervone, D., A.~D’Amour, L.~Bornn, and K.~Goldsberry (2014).
\newblock {POINTWISE:} predicting points and valuing decisions in real time
  with {NBA} optical tracking data.
\newblock In {\em Proceedings of the 8th MIT Sloan Sports Analytics Conference,
  Boston, MA, USA}, Volume~28, pp.\  1--9.

\bibitem[\protect\citeauthoryear{Cervone, D’Amour, Bornn, and
  Goldsberry}{Cervone et~al.}{2016}]{cervone2016multiresolution}
Cervone, D., A.~D’Amour, L.~Bornn, and K.~Goldsberry (2016).
\newblock A multiresolution stochastic process model for predicting basketball
  possession outcomes.
\newblock {\em Journal of the American Statistical Association (Applications
  and Case Studies)\/}~{\em 111\/}(514), 585--599.

\bibitem[\protect\citeauthoryear{Congdon}{Congdon}{2019}]{congdon2019bayesian}
Congdon, P.~D. (2019).
\newblock {\em Bayesian Hierarchical Models: With Applications Using R}.
\newblock Boca Raton, FL: Chapman \& Hall/CRC.

\bibitem[\protect\citeauthoryear{Dixon and Coles}{Dixon and
  Coles}{1997}]{dixon1997modelling}
Dixon, M.~J. and S.~G. Coles (1997).
\newblock Modelling association football scores and inefficiencies in the
  football betting market.
\newblock {\em Journal of the Royal Statistical Society: Series C (Applied
  Statistics)\/}~{\em 46\/}(2), 265--280.

\bibitem[\protect\citeauthoryear{Fernandez and Bornn}{Fernandez and
  Bornn}{2018}]{fernandez2018wide}
Fernandez, J. and L.~Bornn (2018).
\newblock Wide open spaces: A statistical technique for measuring space
  creation in professional soccer.
\newblock In {\em Sloan Sports Analytics Conference}, Volume 2018.

\bibitem[\protect\citeauthoryear{Franks, Miller, Bornn, and Goldsberry}{Franks
  et~al.}{2015}]{franks2015characterizing}
Franks, A., A.~Miller, L.~Bornn, and K.~Goldsberry (2015).
\newblock Characterizing the spatial structure of defensive skill in
  professional basketball.
\newblock {\em The Annals of Applied Statistics\/}~{\em 9\/}(1), 94--121.

\bibitem[\protect\citeauthoryear{Hu, Yang, Xue, and Dey}{Hu
  et~al.}{2023}]{hu2021cjs}
Hu, G., H.-C. Yang, Y.~Xue, and D.~K. Dey (2023).
\newblock Zero-inflated {P}oisson model with clustered regression coefficients:
  Application to heterogeneity learning of field goal attempts of professional
  basketball players.
\newblock {\em Canadian Journal of Statistics\/}.
\newblock To appear.

\bibitem[\protect\citeauthoryear{Karlis and Ntzoufras}{Karlis and
  Ntzoufras}{2003}]{karlis2003analysis}
Karlis, D. and I.~Ntzoufras (2003).
\newblock Analysis of sports data by using bivariate {P}oisson models.
\newblock {\em Journal of the Royal Statistical Society: Series D (The
  Statistician)\/}~{\em 52\/}(3), 381--393.

\bibitem[\protect\citeauthoryear{Miller and Bornn}{Miller and
  Bornn}{2017}]{miller2017possession}
Miller, A.~C. and L.~Bornn (2017).
\newblock Possession sketches: Mapping {NBA} strategies.
\newblock In {\em Proceedings of the 2017 MIT Sloan Sports Analytics
  Conference}, pp.\  1--12.

\bibitem[\protect\citeauthoryear{{Norris}}{{Norris}}{1997}]{Nj97}
{Norris}, J.~R. (1997).
\newblock {\em Markov Chains}.
\newblock Cambridge: Cambridge University Press.

\bibitem[\protect\citeauthoryear{Perry and Wolfe}{Perry and
  Wolfe}{2013}]{Perry2010}
Perry, P.~O. and P.~J. Wolfe (2013).
\newblock Point process modelling for directed interaction networks.
\newblock {\em Journal of the Royal Statistical Society, Series B (Statistical
  Methodology)\/}~{\em 75}, 821--849.

\bibitem[\protect\citeauthoryear{Sandholtz, Mortensen, and Bornn}{Sandholtz
  et~al.}{2020}]{sandholtz2019measuring}
Sandholtz, N., J.~Mortensen, and L.~Bornn (2020).
\newblock Measuring spatial allocative efficiency in basketball.
\newblock {\em Journal of Quantitative Analysis in Sports\/}~{\em 16\/}(4),
  271--289.

\bibitem[\protect\citeauthoryear{Santos-Fernandez, Denti, Mengersen, and
  Mira}{Santos-Fernandez et~al.}{2022}]{santos2022role}
Santos-Fernandez, E., F.~Denti, K.~Mengersen, and A.~Mira (2022).
\newblock The role of intrinsic dimension in high-resolution player tracking
  data—insights in basketball.
\newblock {\em The Annals of Applied Statistics\/}~{\em 16\/}(1), 326--348.

\bibitem[\protect\citeauthoryear{Sisson, Fan, and Tanaka}{Sisson
  et~al.}{2007}]{Sisson1760}
Sisson, S.~A., Y.~Fan, and M.~M. Tanaka (2007).
\newblock Sequential {M}onte {C}arlo without likelihoods.
\newblock {\em Proceedings of the National Academy of Sciences\/}~{\em 104},
  1760--1765.

\bibitem[\protect\citeauthoryear{{Stan Development Team}}{{Stan Development
  Team}}{2023}]{rstan}
{Stan Development Team} (2023).
\newblock {RStan}: the {R} interface to {Stan}.
\newblock R package version 2.21.8.

\bibitem[\protect\citeauthoryear{Stroock}{Stroock}{2014}]{St14}
Stroock, D.~W. (2014).
\newblock {\em An Introduction to Markov Processes\/} (2 ed.).
\newblock Heidelberg: Springer.

\bibitem[\protect\citeauthoryear{Toni, Welch, Strelkowa, Ipsen, and
  Stumpf}{Toni et~al.}{2009}]{Toni09}
Toni, T., D.~Welch, N.~Strelkowa, A.~Ipsen, and M.~P.~H. Stumpf (2009).
\newblock Approximate {B}ayesian computation scheme for parameter inference and
  model selection in dynamical systems.
\newblock {\em Journal of the Royal Society Interface\/}~{\em 6}, 187--202.

\bibitem[\protect\citeauthoryear{Wu and Bornn}{Wu and
  Bornn}{2018}]{wu2018modeling}
Wu, S. and L.~Bornn (2018).
\newblock Modeling offensive player movement in professional basketball.
\newblock {\em The American Statistician\/}~{\em 72\/}(1), 72--79.

\end{thebibliography}

\vskip .65cm

\noindent
Department of Statistics, 
University of Missouri-Columbia
\vskip 2pt
\noindent
E-mail: \href{mailto:grieshopn@missouri.edu}{grieshopn@missouri.edu}
\vskip 2pt

\noindent
Department of Economics, 
University of Missouri-Columbia
\vskip 2pt
\noindent
E-mail: \href{mailto:yong.feng@mail.missouri.edu}{yong.feng@mail.missouri.edu}
\vskip 2pt

\noindent
Department of Biostatistics and Data Science,
The University of Texas Health Science Center at Houston
\vskip 2pt
\noindent
E-mail: \href{mailto:Guanyu.Hu@uth.tmc.edu}{guanyu.hu@uth.tmc.edu}
\vskip 2pt

\noindent
Department of Statistics, 
The Pennsylvania State University
\vskip 2pt
\noindent
E-mail: \href{mailto:michael.schweinberger@psu.edu}{michael.schweinberger@psu.edu}
\vskip 2pt


\newpage

\setcounter{page}{1}

\begin{appendix}

\fontsize{12}{14pt plus.8pt minus .6pt}\selectfont \vspace{0.8pc}
\centerline{\large\bf SUPPLEMENT}
\vspace{2pt} 
\centerline{\large\bf A CONTINUOUS-TIME STOCHASTIC PROCESS FOR}
\vspace{2pt} 
\centerline{\large\bf HIGH-RESOLUTION NETWORK DATA IN SPORTS}
\vspace{.4cm} 
\centerline{Nicholas Grieshop$^1$, Yong Feng$^2$, Guanyu Hu$^3$ and Michael Schweinberger$^4$}
\vspace{.4cm} 
\centerline{\it 1 Department of Statistics, University of Missouri-Columbia}
\vspace{.4cm} 
\centerline{\it 2 Department of Economics, University of Missouri-Columbia}
\begin{center}
{\it 3 Department of Biostatistics and Data Science,
The University of Texas Health Science Center at Houston}
\end{center}
\centerline{\it 4 Department of Statistics, The Pennsylvania State University}
\vspace{.55cm} \fontsize{9}{11.5pt plus.8pt minus.6pt}\selectfont

\thispagestyle{empty}



\noindent\normalfont
{Supplement \ref{sec:plots}: Figures}\dotfill\pageref{sec:plots}\s
\\
{Supplement \ref{sec:abbreviations}: Abbreviations}\dotfill\pageref{sec:abbreviations}\s
\\
{Supplement \ref{sec:descriptives}: Descriptive statistics}\dotfill\pageref{sec:descriptives}\s
\\
{Supplement \ref{sec:summaries}: Posterior summaries}\dotfill\pageref{sec:summaries}\s
\\
{Supplement \ref{app:Sensitvity}: Posterior sensitivity checks}\dotfill\pageref{app:Sensitvity}\s
\\
{Supplement \ref{sec:posterior.predictive.checks}: Posterior predictive checks}\dotfill\pageref{sec:posterior.predictive.checks}\s
\\
{Supplement \ref{sec:simulation.results}: Simulation results}\dotfill\pageref{sec:simulation.results}\s

\clearpage

\section{Figures}
\label{sec:plots}

\begin{figure}[htp]
    \centering
    \begin{minipage}{0.5\linewidth}
        \centering
        \includegraphics[width=1\linewidth]{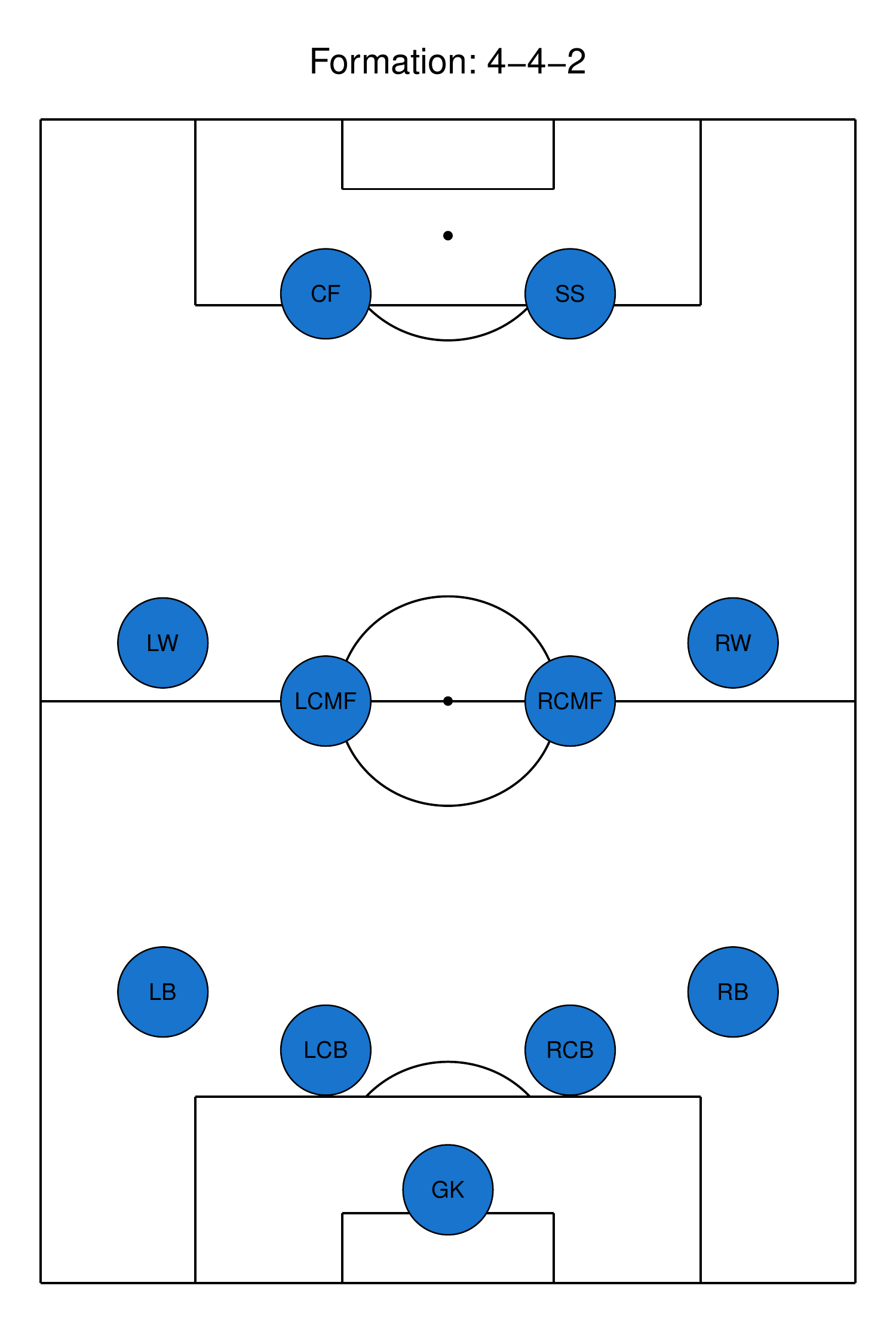}
    \end{minipage}%
    \begin{minipage}{0.5\linewidth}
        \centering
        \includegraphics[width=1\linewidth]{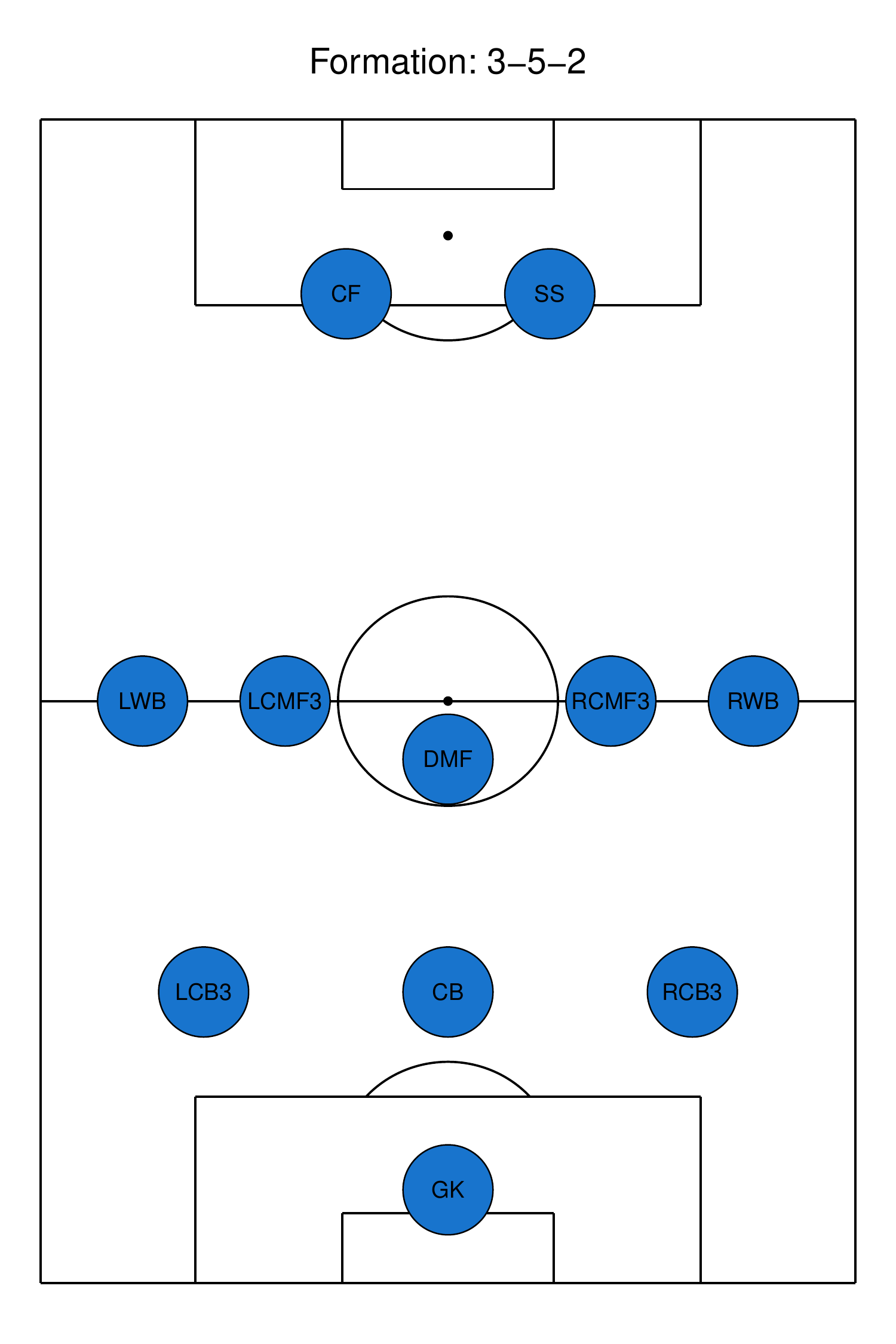}
    \end{minipage}
    \caption{Two popular formations of soccer teams, known as 4-4-2 and 3-5-2. The abbreviations of player positions are detailed in Supplement \ref{sec:abbreviations}.
    \label{fig:formation}
}
\end{figure}

\begin{figure}
\centering
	\subfigure[Juventus Turin (4-4-2)]{\includegraphics[width=0.485\textwidth]{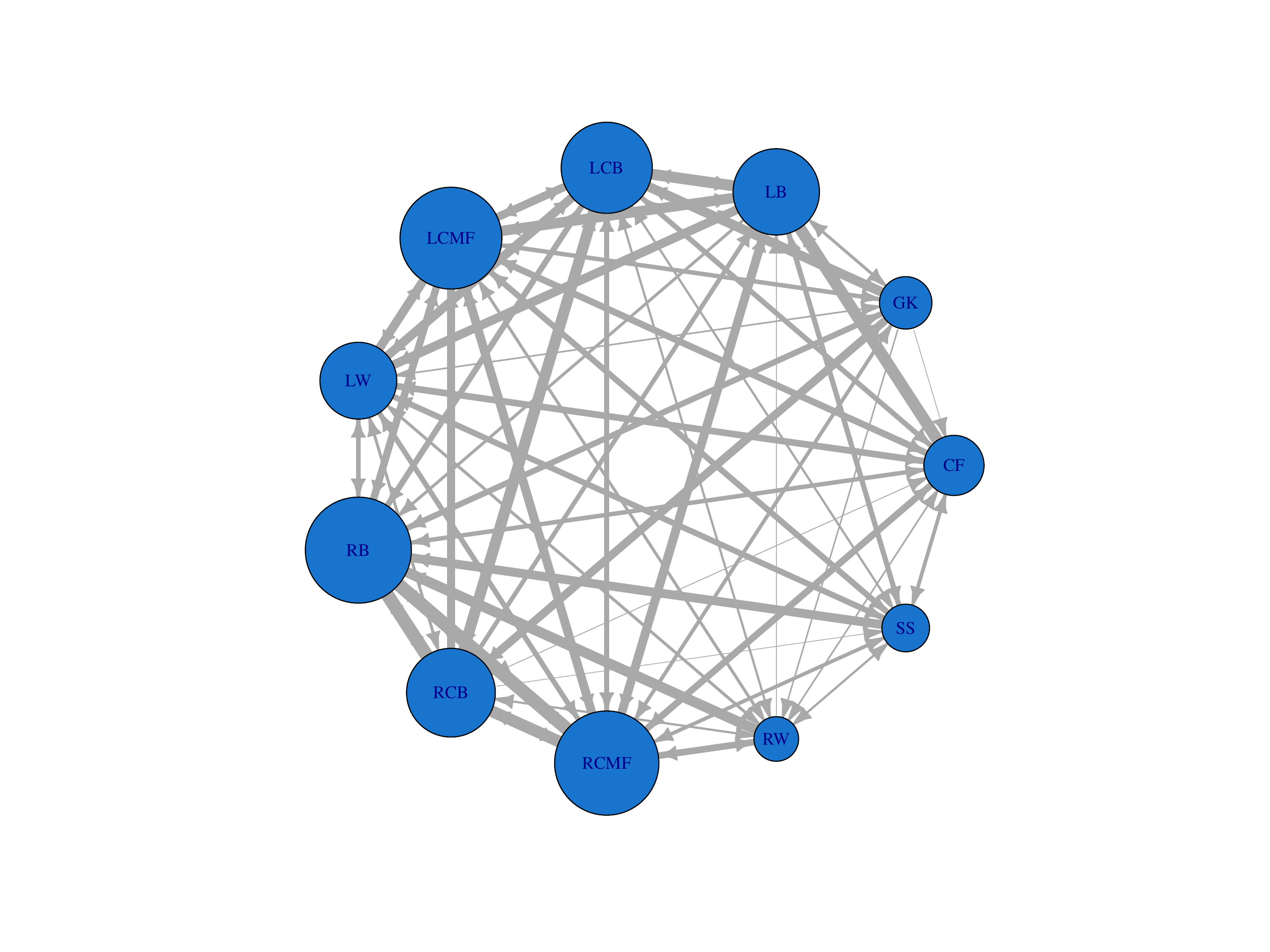}
 \label{fig:network-juventus}}
\subfigure[Inter Milan (3-5-2)]{\includegraphics[width=0.485\textwidth]{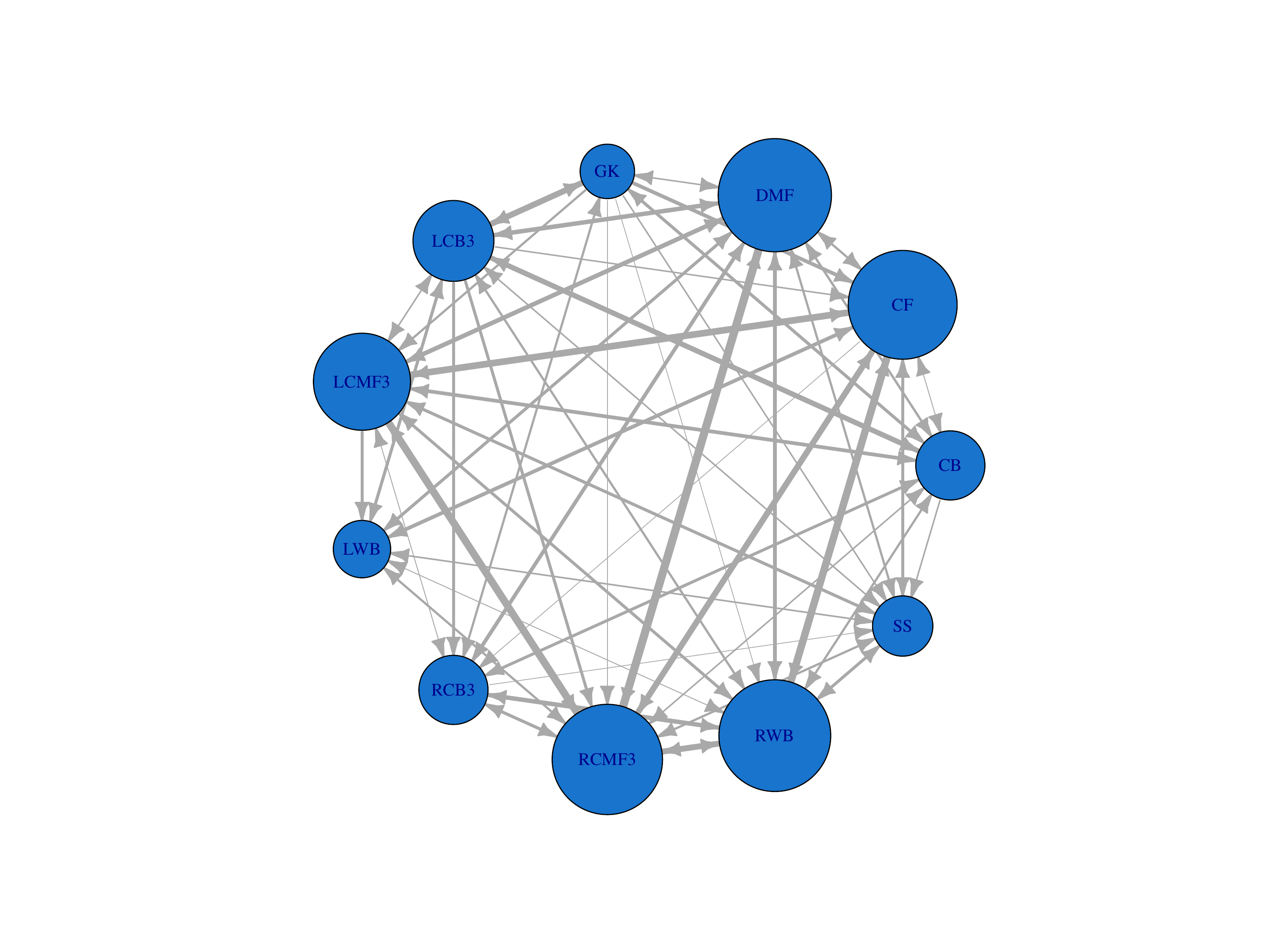}
\label{fig:network-inter}}
\caption{The numbers of passes between the positions of (a) Juventus Turin (with 4-4-2 formation) and (b) Inter Milan (with 3-5-2 formation).
These data are based on the home games of (a) Juventus Turin versus AC Milan and (b) Inter Milan versus AC Milan in 2020/21.
The 4-4-2 and 3-5-2 formations are shown in Figure \ref{fig:formation} in Supplement \ref{sec:plots}.
The sizes of the positions are proportional to the number of passes,
while the widths of the edges are proportional to the number of passes between the positions.}
    \label{fig:network-comb}
\end{figure}

\begin{figure}
    \centering
    \begin{minipage}{0.45\linewidth}
        \centering
        \includegraphics[width=1\linewidth]{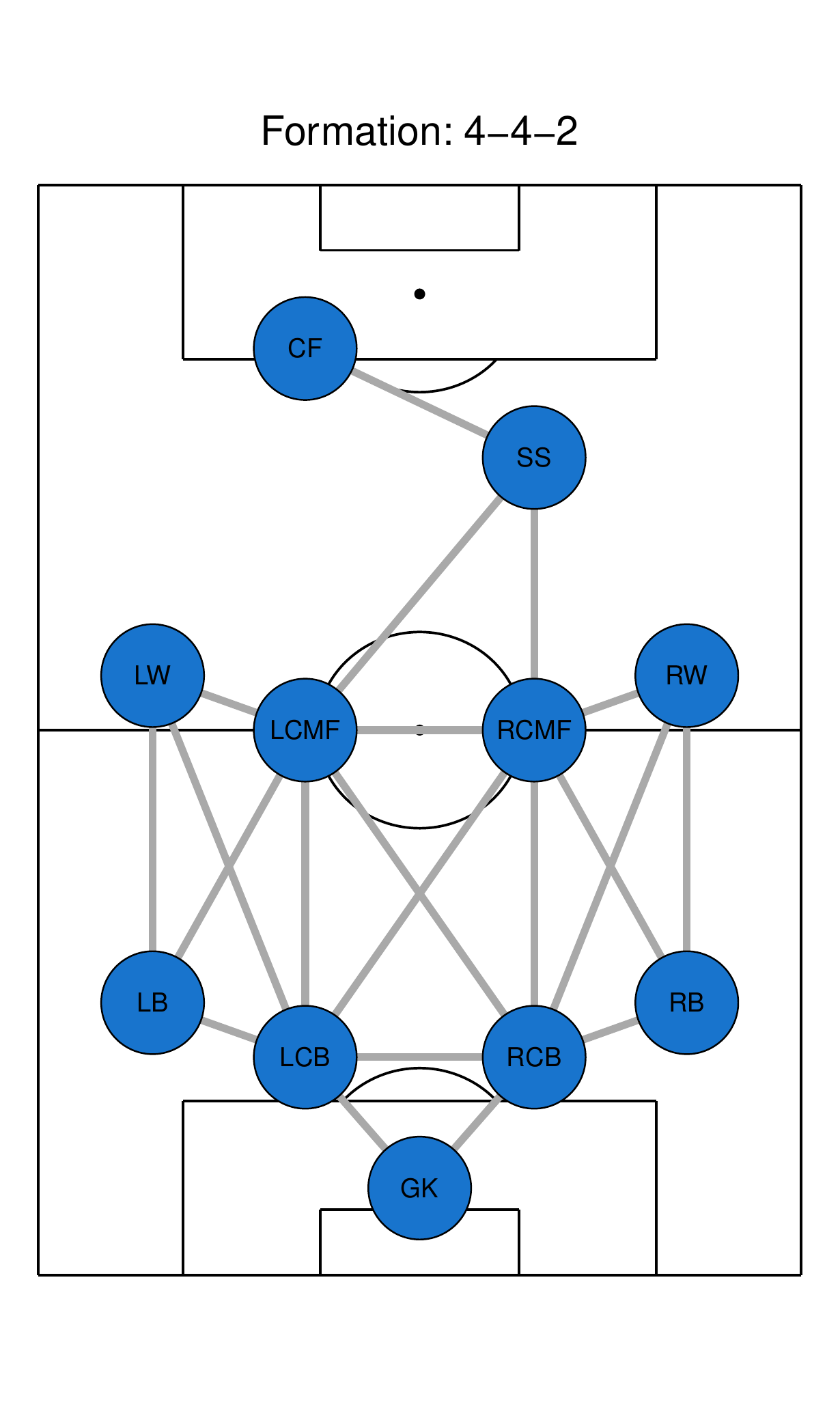}
    \end{minipage}%
    \begin{minipage}{0.45\linewidth}
        \centering
        \includegraphics[width=1\linewidth]{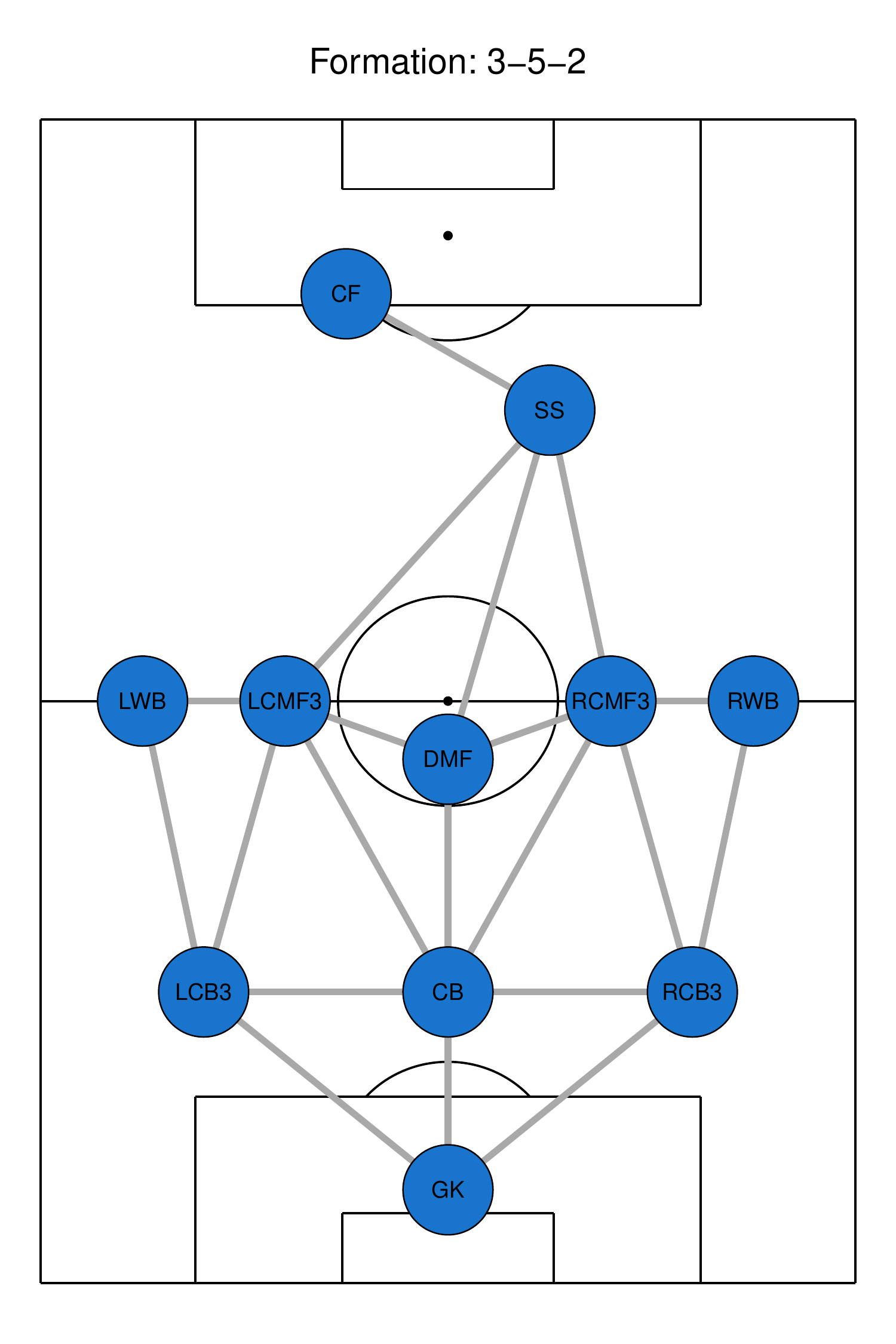}
    \end{minipage}
    \caption{\label{nngraph2} 
    The nearest-neighbor graph, 
which connects pairs of positions that are considered to be nearest neighbors on the \textcolor{black}{field}. The graph distance between a pair of positions is the length of the shortest path between them. The abbreviations of player positions are detailed in Supplement \ref{sec:abbreviations}.
}
\end{figure}

\clearpage

\section{Abbreviations}
\label{sec:abbreviations}

\begin{table}[!htbp]
\centering
\caption{
\label{tab:positionPass}Abbreviations of player positions.
}\s
\begin{tabular}{@{}ll}
\hline
Player position                         & Abbreviation \\
\hline
Center Back                             & CB                   \\
Right Center Back                       & RCB                \\
Left Center Back                        & LCB                   \\
Left Defensive Midfielder               & LDMF                   \\
Right Defensive Midfielder              & RDMF                  \\
Right Center Back (3 at the back)       & RCB3                  \\
Goalkeeper                              & GK                 \\
Defensive Midfielder                    & DMF                  \\
Left Center Midfielder                  & LCMF                  \\
Left Center Back (3 at the back)        & LCB3                  \\
Right Center Midfielder                 & RCMF                  \\
Left Center Midfielder (3 at the back)  & LCMF3                   \\
Right Back                              & RB                   \\
Left Back                               & LB                  \\
Attacking Midfielder                    & AMF                   \\
Right Center Midfielder (3 at the back) & RCMF3                  \\
Left Attacking Midfielder               & LAMF                 \\
Left Wing Forward                       & LWF                  \\
Right Wing Forward                      & RWF                  \\
Left Wing                               & LW                   \\
Right Attacking Midfielder              & RAMF                   \\
Right Wing Back                         & RWB                   \\
Second Striker                          & SS                   \\
Right Wing                              & RW                  \\
Left Wing Back                          & LWB                  \\
Striker                          & CF                   \\
Left Back (5 at the back)               & LB5                 \\
Right Back (5 at the back)              & RB5         \\
\hline
\end{tabular}

\end{table}

\clearpage

\section{Descriptive statistics}
\label{sec:descriptives}

\begin{table}[!htbp]
    \centering
        \caption{Successful proportion of passes by teams during the 2020/21 season of the Italian premier football league.
}\s

    \begin{tabular}{lc c c}
    \hline
          & \multicolumn{3}{c}{Successful proportion of passes}  \\
        Team & Total & First half & Second half \\
        \hline
Sassuolo& $89.47\%$ & $89.97\%$ & $88.92\%$ \\
Juventus Turin &$88.95\%$ & $89.56\%$ & $88.27\%$ \\
Inter Milan& $88.55\%$ & $88.69\%$ & $88.40\%$ \\
Napoli& $88.10\%$ & $88.67\%$ & $87.46\%$ \\
Roma& $86.44\%$ & $87.17\%$ & $85.65\%$ \\
Atalanta& $86.33\%$ & $86.33\%$ & $86.32\%$ \\
Milan & $86.22\%$ & $86.59\%$ & $85.82\%$ \\
Lazio & $85.91\%$ & $86.36\%$ & $85.43\%$ \\
Parma & $85.07\%$ & $85.86\%$ & $84.27\%$ \\
Udinese & $84.79\%$ & $85.57\%$ & $84.01\%$ \\
Torino & $84.68\%$ & $85.75\%$ & $83.53\%$ \\
Fiorentina & $84.50\%$ & $85.37\%$ & $83.64\%$ \\
Bologna & $84.45\%$ & $84.71\%$ & $84.17\%$ \\
Spezia & $84.39\%$ & $84.84\%$ & $83.91\%$ \\
Crotone & $84.28\%$ & $84.53\%$ & $84.01\%$ \\
Cagliari & $83.50\%$ & $84.41\%$ & $82.55\%$ \\
Genoa & $83.20\%$ & $84.13\%$ & $82.27\%$ \\
Benevento & $81.70\%$ & $81.52\%$ & $81.87\%$ \\
Hellas Verona & $81.39\%$ & $82.14\%$ & $80.58\%$ \\
Sampdoria & $81.35\%$ & $81.97\%$ & $80.70\%$ \\
   \hline
    \end{tabular}
    \label{tab:totalPass}
\end{table}

\begin{table}[!htbp]
\centering
\caption{Number of passes and successful passes of Juventus Turin during the 2020/21 season,
by formation.}\s

\begin{tabular}{@{}lll@{}}
   \hline
Formation & Number of passes & Proportion of successful passes \\ \midrule
4-4-2          & 15832       & 89.30\% \\
4-4-1-1        & 1529        & 88.69\% \\
4-4-1          & 744         & 88.04\% \\
3-4-2-1        & 737         & 91.18\% \\
4-2-3-1        & 591         & 87.14\% \\
3-5-2          & 455         & 81.98\% \\
3-4-1-2        & 261         & 88.89\% \\
3-5-1-1        & 147         & 87.76\% \\
3-4-3          & 114         & 81.58\% \\
4-3-1-2        & 95          & 84.21\% \\
4-3-2          & 54          & 87.04\% \\
4-5-1          & 39          & 84.62\% \\
5-3-1          & 3           & 66.67\% \\
5-4-1          & 1           & 0.00\%  \\ \bottomrule
\end{tabular}

\label{tab:formation_juventus}
\end{table}

\begin{table}[!htbp]
\centering
\caption{Number of passes and successful passes of Juventus Turin during the 2020/21 season based on the 4-4-2 formation,
by position and player.
Danilo refers to the player Danilo Luiz da Silva.}\s
\scriptsize
\begin{tabular}{@{}llll@{}}
\hline
Position & Player      & Number of passes & Proportion of successful passes  \\ \midrule
CF               & C. Ronaldo & 741         & 81.38\%  \\
               & A. Morata     & 197         & 76.65\%  \\
               & P. Dybala         & 5           & 80.00\%  \\
GK               & W. Szczesny       & 559         & 95.53\%  \\
              & G. Buffon         & 173         & 91.33\%  \\
               & C. Pinsoglio      & 8           & 87.50\%  \\
LB               & A. Sandro       & 769         & 88.43\%  \\
              & Danilo           & 535         & 91.78\%  \\
              & G. Frabotta       & 239         & 87.45\%  \\
               & F. Bernardeschi   & 152         & 88.16\%  \\
               & J. Cuadrado       & 26          & 92.31\%  \\
LCB              & G. Chiellini      & 786         & 92.75\%  \\
            & M. de Ligt        & 518         & 94.59\%  \\
             & L. Bonucci        & 399         & 90.23\%  \\
             & Danilo            & 91          & 95.60\%  \\
             & M. Demiral        & 77          & 98.70\%  \\
              & A. Sandro       & 72          & 90.28\%  \\
LCMF             & A. Rabiot         & 765         & 92.16\%  \\
            & R. Bentancur      & 478         & 91.84\%  \\
             & A. Melo            & 368         & 95.65\%  \\
            & W. McKennie       & 100         & 87.00\%  \\
            & N. Fagioli        & 17          & 100.00\% \\
           & A. Ramsey         & 9           & 100.00\% \\
LW               & A. Ramsey         & 400         & 89.75\%  \\
            & F. Chiesa         & 345         & 79.13\%  \\
            & F. Bernardeschi   & 153         & 81.70\%  \\
             & W. McKennie       & 140         & 85.00\%  \\
             & D. Kulusevski     & 51          & 80.39\%  \\
             & G. Frabotta       & 23          & 78.26\%  \\
              & F. Correia     & 8           & 87.50\%  \\
           & A. Rabiot         & 4           & 100.00\% \\
RB               & J. Cuadrado       & 1010        & 85.15\%  \\
             & Danilo            & 883         & 88.34\%  \\
             & M. Demiral        & 5           & 80.00\%  \\
RCB              & M. de Ligt        & 798         & 95.49\%  \\
              & L. Bonucci        & 627         & 92.50\%  \\
              & M. Demiral        & 328         & 96.95\%  \\
              & Danilo          & 35          & 97.14\%  \\
              & R. Drăgușin       & 2           & 50.00\%  \\
RCMF             & R. Bentancur      & 835         & 90.30\%  \\
             & A. Melo            & 501         & 94.81\%  \\
             & A. Rabiot         & 267         & 92.51\%  \\
             & Danilo            & 145         & 92.41\%  \\
             & W. McKennie       & 129         & 93.02\%  \\
             & M. Portanova      & 5           & 100.00\% \\
             & A. Ramsey         & 3           & 100.00\% \\
RW               & D. Kulusevski     & 386         & 80.05\%  \\
               & F. Chiesa         & 243         & 80.25\%  \\
               & W. McKennie       & 164         & 85.37\%  \\
              & J. Cuadrado       & 159         & 87.42\%  \\
              & A. Ramsey         & 82          & 85.37\%  \\
              & F. Bernardeschi   & 19          & 78.95\%  \\
              & P. Dybala         & 4           & 100.00\% \\
             & D. Costa     & 3           & 33.33\%  \\
                   & G. Vrioni         & 2           & 100.00\% \\
SS               & P. Dybala         & 495         & 87.88\%  \\
               & Á. Morata     & 305         & 80.33\%  \\
               & D. Kulusevski     & 113         & 80.53\%  \\
               & C. Ronaldo & 73          & 76.71\%  \\
               & F. Chiesa         & 1           & 0.00\%   \\ \bottomrule
\end{tabular}

\label{tab:juventus_position_comp}
\end{table}

\begin{table}[!htbp]
\small
\centering
\caption{Number of passes and successful passes of Juventus Turin based on the 4-4-2 formation during the 2020/21 season, by player and position.}\s 

\label{app:juventus_famous}
\begin{tabular}{@{}llll@{}}
\toprule
Player       & Position & Number of passes & Proportion of successful passes \\ \midrule
C.\ Ronaldo & CF              & 908              & 81.28\%         \\
 & SS              & 80               & 76.25\%         \\
 & AMF             & 18               & 88.89\%         \\
D.\ Kulusevski     & RW              & 386              & 80.05\%         \\
     & SS              & 136              & 80.88\%         \\
     & LW              & 87               & 79.31\%         \\
     & AMF             & 67               & 83.58\%         \\
     & RAMF            & 26               & 80.77\%         \\
     & RWF             & 12               & 75.00\%         \\
     & CF              & 9                & 66.67\%         \\
     & RCMF            & 4                & 75.00\%         \\
     & LCMF3           & 3                & 33.33\%         \\
F.\ Chiesa         & LW              & 367              & 79.56\%         \\
         & RW              & 294              & 80.95\%         \\
         & RWB             & 44               & 61.36\%         \\
         & LAMF            & 22               & 72.73\%         \\
         & LWB             & 4                & 50.00\%         \\
         & RCMF3           & 4                & 50.00\%         \\
         & RAMF            & 2                & 50.00\%         \\
         & SS              & 1                & 0.00\%          \\
M.\ de Ligt        & RCB             & 964              & 95.64\%         \\
        & LCB             & 528              & 94.51\%         \\
        & RCB3            & 64               & 92.19\%         \\
        & CB              & 39               & 92.31\%         \\
P.\ Dybala         & SS              & 512              & 87.70\%         \\
         & AMF             & 81               & 91.36\%         \\
         & CF              & 10               & 90.00\%         \\
         & LW              & 7                & 85.71\%         \\
         & RW              & 4                & 100.00\%        \\
R.\ Bentancur      & RCMF            & 1042             & 90.60\%         \\
      & LCMF            & 548              & 91.97\%         \\
      & DMF             & 64               & 85.94\%         \\
      & RCMF3           & 20               & 90.00\%         \\
      & LCMF3           & 9                & 77.78\%         \\ \bottomrule
\end{tabular}
\end{table}

\begin{table}[!htbp]
\centering
\caption{Number of passes and successful passes of Inter Milan during the 2020/21 season,
by formation.\s
}\s
\begin{tabular}{@{}lll@{}}
\toprule
Formation & Number of passes & Proportion of successful passes \\ \midrule
3-5-2          & 13564       & 88.85\% \\
3-4-1-2        & 3329        & 88.80\% \\
5-3-2          & 1098        & 85.70\% \\
3-4-3          & 485         & 86.19\% \\
4-3-1-2        & 262         & 93.51\% \\
5-4-1          & 172         & 78.49\% \\
3-4-2-1        & 110         & 87.27\% \\
3-4-2          & 60          & 90.00\% \\
3-5-1-1        & 57          & 89.47\% \\
4-4-1-1        & 28          & 85.71\% \\
4-3-2          & 14          & 85.71\% \\ \bottomrule
\end{tabular}

\label{tab:formation_inter}
\end{table}

\begin{table}[!htbp]
\centering
\caption{Number of passes and successful passes of Inter Milan during the 2020/21 season based on the 3-5-2 formation,
by position and player.}\s
\scriptsize
\begin{tabular}{@{}llll@{}}
\hline
Position & Player   & Number of passes & Proportion of successful passes \\ \midrule
CB               & S. de Vrij     & 1371        & 96.21\%  \\
               & A. Ranocchia   & 296         & 95.95\%  \\
CF               & R. Lukaku      & 305         & 76.72\%  \\
               & A. Sánchez     & 174         & 81.61\%  \\
               & L. Martínez    & 136         & 80.88\%  \\
               & I. Perišić     & 5           & 60.00\%  \\
               & A. Pinamonti   & 1           & 100.00\% \\
DMF              & M. Brozović    & 1518        & 91.77\%  \\
              & C. Eriksen     & 204         & 88.73\%  \\
             & N. Barella     & 60          & 88.33\%  \\
              & A. Vidal       & 28          & 89.29\%  \\
              & R. Gagliardini & 23          & 91.30\%  \\
GK               & S. Handanovič  & 557         & 90.84\%  \\
             & I. Radu        & 28          & 100.00\% \\
             & D. Padelli     & 9           & 100.00\% \\
LCB3             & A. Bastoni     & 1578        & 92.27\%  \\
             & M. Škriniar    & 86          & 94.19\%  \\
             & A. Kolarov     & 43          & 86.05\%  \\
             & M. Darmian     & 25          & 100.00\% \\
LCMF3            & C. Eriksen     & 419         & 88.78\%  \\
            & R. Gagliardini & 347         & 90.78\%  \\
            & A. Vidal       & 245         & 89.39\%  \\
            & S. Sensi       & 204         & 88.73\%  \\
            & N. Barella     & 102         & 90.20\%  \\
LWB              & I. Perišić     & 390         & 77.18\%  \\
              & A. Young       & 354         & 82.20\%  \\
             & M. Darmian     & 97          & 81.44\%  \\
              & D. D'Ambrosio  & 5           & 80.00\%  \\
RCB3             & M. Škriniar    & 1526        & 94.82\%  \\
             & D. D'Ambrosio  & 250         & 93.60\%  \\
             & S. de Vrij     & 21          & 95.24\%  \\
RCMF3            & N. Barella     & 1114        & 84.11\%  \\
            & A. Vidal       & 157         & 84.71\%  \\
            & M. Vecino      & 97          & 86.60\%  \\
            & S. Sensi       & 30          & 86.67\%  \\
            & C. Eriksen     & 25          & 92.00\%  \\
            & R. Gagliardini & 12          & 83.33\%  \\
           & R. Nainggolan  & 2           & 100.00\% \\
RWB              & A. Hakimi      & 950         & 82.32\%  \\
             & M. Darmian     & 166         & 83.13\%  \\
              & A. Young       & 12          & 100.00\% \\
              & D. D'Ambrosio  & 5           & 100.00\% \\
SS               & L. Martínez    & 228         & 71.49\%  \\
              & A. Sánchez     & 165         & 80.00\%  \\
              & R. Lukaku      & 164         & 71.95\%  \\
              & A. Pinamonti   & 26          & 76.92\%  \\ \bottomrule
\end{tabular}

\label{tab:inter_position_comp}
\end{table}

\begin{table}[!htbp]
\small
\centering
\caption{Number of passes and successful passes of Inter Milan during the 2020/21 season by famous players in different positions.}\s

\label{app:inter_famous}
\begin{tabular}{@{}llll@{}}
\toprule
Player & Position & Number of passes & Proportion of successful passes \\ \midrule
C. Eriksen  & LCMF3           & 507              & 88.76\%         \\
  & DMF             & 212              & 89.15\%         \\
  & AMF             & 141              & 82.98\%         \\
  & RCMF3           & 25               & 92.00\%         \\
  & LCMF            & 8                & 75.00\%         \\
  & RCMF            & 4                & 100.00\%        \\
  & SS              & 4                & 50.00\%         \\
L. Martínez & SS              & 313              & 69.33\%         \\
 & CF              & 195              & 78.97\%         \\
 & LWF             & 19               & 73.68\%         \\
 & LW              & 2                & 100.00\%        \\
M. Brozović & DMF             & 1661             & 91.75\%         \\
 & RCMF            & 272              & 87.50\%         \\
 & LCMF            & 108              & 89.81\%         \\
M. Škriniar & RCB3            & 1950             & 94.82\%         \\
 & LCB3            & 87               & 93.10\%         \\
 & RCB             & 18               & 94.44\%         \\
N. Barella  & RCMF3           & 1245             & 84.58\%         \\
  & RCMF            & 176              & 91.48\%         \\
  & LCMF            & 140              & 88.57\%         \\
  & LCMF3           & 114              & 90.35\%         \\
  & DMF             & 80               & 88.75\%         \\
  & AMF             & 70               & 88.57\%         \\
  & RW              & 8                & 75.00\%         \\
  & LWF             & 4                & 100.00\%        \\
R. Lukaku   & CF              & 462              & 76.41\%         \\
   & SS              & 244              & 75.00\%         \\ \bottomrule
\end{tabular}
\end{table}

\clearpage

\section{Posterior summaries}
\label{sec:summaries}

\begin{center}
\begin{table}[]
\caption{{Posterior summaries for Fiorentina, Crotone, and Inter Milan (with 3-5-2 formation): M refers to posterior medians and CI refers to 95\% posterior credible intervals.
}}
\s
    \centering
    \begin{tabular}{lrcrcrc}
        & \multicolumn{2}{c}{\bf Fiorentina} &  \multicolumn{2}{c}{\bf Crotone} &  \multicolumn{2}{c}{\bf Inter Milan} \s
        \\
          & M & CI & M & CI & M & CI\\
        \midrule
        \multicolumn{6}{l}{\bf Successful passes $\{S_{i_m} = 1\}$:}
        \\
        Intercept & 2.93 & (2.47,\;3.39) & 3.27 & (2.85,\;3.68) & 3.34 & (2.82,\;3.86)\\
        Length of pass & 0.00 & (-0.01,\;0.00) & 0.00 & (-0.01,\;0.00) & 0.00 & (0.00,\;0.01)\\
        Forward pass & -0.57 & (-0.74,\;-0.40) & -0.88 & (-1.07,\;-0.70) & -0.84 & (-0.99,\;-0.70)\\
        Start: half & 0.17 & (-0.02,\;0.36) & -0.03 & (-0.23,\;0.17) & 0.29 & (0.12,\;0.46)\\
        End: third & -0.67 & (-0.85,\;-0.49) & -0.64 & (-0.84,\;-0.45) & -0.79 & (-0.96,\;-0.62)\\
        Air pass & -1.76 & (-1.93,\;-1.59) & -1.90 & (-2.07,\;-1.73) & -1.84 & (-1.98,\;-1.70)\\
        Winning & -0.13 & (-0.30,\;0.04) & -0.25 & (-0.46,\;-0.04) & -0.13 & (-0.26,\;0.00)\\
        Losing & -0.01 & (-0.17,\;0.15) & -0.11 & (-0.26,\;0.04) & 0.02 & (-0.16,\;0.21)\\
        \hline
        \multicolumn{6}{l}{\bf Passes $\{i_m \to j_m\}$ given $\{S_{i_m} = 1\}$:}\\
        Graph distance & -0.69 & (-0.73,\;-0.65) & -0.70 & (-0.74,\;-0.65) & -0.98 & (-1.02,\;-0.95)\\
        Pass received 
        & 0.00 
        & (-2.3e-3, 2.0e-3) 
        & 0.00 
        & (-1.7e-3, 3.2e-4) 
        & 0.00 
        & (-1.6e-3, 8.6e-5) \\
        \hline
        \multicolumn{6}{l}{\bf Holding times $h_m$:}\\
        GK & -3.23 & (-3.34,\;-3.13) & -2.86 & (-2.95,\;-2.77) & -2.98 & (-3.06,\;-2.90)\\
        LCB & -2.62 & (-2.68,\;-2.56) & -2.77 & (-2.83,\;-2.71) & -2.32 & (-2.37,\;-2.27)\\
        CB & -2.62 & (-2.70,\;-2.55) & -2.70 & (-2.76,\;-2.64) & -2.39 & (-2.44,\;-2.34)\\
        RCB & -2.86 & (-2.92,\;-2.79) & -2.44 & (-2.51,\;-2.37) & -2.34 & (-2.39,\;-2.30)\\
        LWB & -2.33 & (-2.40,\;-2.26) & -2.61 & (-2.69,\;-2.53) & -2.27 & (-2.34,\;-2.20)\\
        LCMF & -2.51 & (-2.59,\;-2.43) & -2.50 & (-2.57,\;-2.42) & -2.07 & (-2.13,\;-2.01)\\
        DMF & -2.62 & (-2.68,\;-2.55) & -2.42 & (-2.49,\;-2.36) & -2.13 & (-2.18,\;-2.08)\\
        RCMF & -2.34 & (-2.41,\;-2.26) & -2.62 & (-2.70,\;-2.54) & -2.21 & (-2.26,\;-2.16)\\
        RWB & -2.48 & (-2.56,\;-2.40) & -2.29 & (-2.37,\;-2.20) & -2.01 & (-2.07,\;-1.95)\\
        SS & -2.63 & (-2.72,\;-2.54) & -2.37 & (-2.46,\;-2.28) & -2.11 & (-2.19,\;-2.03)\\
        CF & -2.62 & (-2.71,\;-2.54) & -2.98 & (-3.08,\;-2.88) & -2.30 & (-2.38,\;-2.22)\\
        Winning & -0.42 & (-0.47,\;-0.36) & -0.37 & (-0.44,\;-0.30) & -0.36 & (-0.40,\;-0.33)\\
        Losing & 0.05 & (0.00,\;0.10) & -0.01 & (-0.05,\;0.04) & -0.08 & (-0.13,\;-0.03)\\
        \hline
         \multicolumn{6}{l}{\bf Random effects:}\\
        Correlation & -0.36 & (-0.83,\;0.12) & -0.25 & (-0.75,\;0.25) & -0.03 & (-0.53,\;0.47)\\
        SD: success & 0.58 & (0.31,\;0.86) & 0.62 & (0.34,\;0.89) & 0.80 & (0.44,\;1.17)\\
        SD: pass & 0.51 & (0.28,\;0.74) & 0.24 & (0.13,\;0.36) & 0.47 & (0.25,\;0.69)\\
        \bottomrule
        \end{tabular}
    \label{tab:appRed}
\end{table}
\end{center}

\begin{center}
\begin{table}[]
\caption{{Posterior summaries for Juventus Turin (with 4-4-2 formation):
M refers to posterior medians and CI refers to 95\% posterior credible intervals.}}
\s

    \centering
    \makebox[\linewidth]{
        \begin{tabular}{lrc}
        & \multicolumn{2}{c}{\bf Juventus Turin} \\
          & M & CI\\
        \midrule
        \multicolumn{3}{l}{\bf Successful passes $\{S_{i_m} = 1\}$:}\\
        Intercept  & 3.36 & (2.90,\;3.81)\\
        Length of pass  & 0.00 & (-0.01,\;0.00)\\
        Forward pass & -0.61 & (-0.75,\;-0.47)\\
        Start: half & 0.26 & (0.10,\;0.42)\\
        End: third & -0.92 & (-1.07,\;-0.76)\\
        Air pass & -2.04 & (-2.18,\;-1.89)\\
        Winning & -0.04 & (-0.17,\;0.09)\\
        Losing & 0.04 & (-0.13,\;0.20)\\
        \hline
         \multicolumn{3}{l}{\bf Passes $\{i_m \to j_m\}$ given $\{S_{i_m} = 1\}$:}\\
        Graph distance & -0.70 & (-0.73,\;-0.67)\\
        Pass received  
        & 0.00 
        & (-1.6e-3,\, 8.6e-05)\\
        \hline
        \multicolumn{3}{l}{\bf Holding times $h_m$:}\\
        GK & -2.80 & (-2.88,\;-2.72)\\
        LB & -2.08 & (-2.13,\;-2.03)\\
        LCB & -2.38 & (-2.42,\;-2.33)\\
        RCB & -2.37 & (-2.41,\;-2.32)\\
        RB & -2.09 & (-2.14,\;-2.05)\\
        LW & -2.06 & (-2.12,\;-2.00)\\
        LCMF & -2.19 & (-2.24,\;-2.14)\\
        RCMF & -2.26 & (-2.30,\;-2.21)\\
        RW & -2.17 & (-2.23,\;-2.11)\\
        SS & -1.81 & (-1.87,\;-1.74)\\
        CF & -1.94 & (-2.01,\;-1.87)\\
        Winning & -0.20 & (-0.23,\;-0.16)\\
        Losing & -0.15 & (-0.19,\;-0.10)\\
        \hline
        \multicolumn{3}{l}{\bf Random effects:}\\
        Correlation & -0.33 & (-0.82,\;0.15)\\
        SD: success & 0.69 & (0.38,\;0.99)\\
        SD: pass & 0.44 & (0.24,\;0.64)\\
        \bottomrule
        \end{tabular}
        } 
    \label{tab:appRed_juventus}
\end{table}
\end{center}

\clearpage

\section{Posterior sensitivity checks}
\label{app:Sensitvity}

\begin{center}
\begin{table}[!htbp]
\caption{\textcolor{blue}{Posterior summaries for Fiorentina, Crotone, and Inter Milan (with 3-5-2 formation) under Prior 1 described in Section \ref{sec:sensitivity}: 
M refers to posterior medians and CI refers to 95\% posterior credible intervals.}}
\s
    \centering
    \footnotesize
    \begin{tabular}{lrlrlrl}
        & \multicolumn{2}{c}{\bf Fiorentina} &  \multicolumn{2}{c}{\bf Crotone} &  \multicolumn{2}{c}{\bf Inter Milan} \s
        \\
          & M & CI & M & CI & M & CI\\
        \midrule
        \multicolumn{6}{l}{\bf Successful passes $\{S_{i_m} = 1\}$:}
        \\
        Intercept & 2.94 & (2.56,\;3.31) & 3.23 & (2.82,\;3.64) & 3.24 & (2.74,\;3.75)\\
        Length of pass  & 0.00 & (-0.01,\;0.00) & 0.00 & (-0.01,\;0.00) & 0.00 & (-0.01,\;0.00)\\
        Forward pass & -0.56 & (-0.73,\;-0.4) & -0.88 & (-1.06,\;-0.71) & -0.84 & (-0.99,\;-0.68)\\
        Start: half & 0.15 & (-0.04,\;0.35) & -0.02 & (-0.22,\;0.17) & 0.29 & (0.11,\;0.47)\\
        End: third &  -0.68 & (-0.87,\;-0.49) & -0.62 & (-0.82,\;-0.43) & -0.79 & (-0.97,\;-0.62)\\
        Air pass & -1.76 & (-1.93,\;-1.59) & -1.90 & (-2.07,\;-1.73) & -1.84 & (-1.99,\;-1.7)\\
        Winning & -0.15 & (-0.32,\;0.03) & -0.23 & (-0.45,\;-0.02) & -0.12 & (-0.25,\;0.02)\\
        Losing & -0.01 & (-0.19,\;0.16) & -0.12 & (-0.27,\;0.02) & 0.01 & (-0.17,\;0.19)\\
        \hline
        \multicolumn{6}{l}{\bf Passes $\{i_m \to j_m\}$ given $\{S_{i_m} = 1\}$:}\\
        Graph distance & -0.69 & (-0.73,\;-0.65) & -0.70 & (-0.74,\;-0.65) & -0.99 & (-1.02,\;-0.95)\\
        Pass received & 0.00 & (-2.09e-3,\;1.92e-3) & 0.00 & (-1.68e-3,\;3.13e-4) & 0.00 & (-1.52e-3,\;4.37e-05)\\\\
        \hline
        \multicolumn{6}{l}{\bf Holding times $h_m$:}\\
        GK & -3.23 & (-3.33,\;-3.13) & -2.86 & (-2.94,\;-2.77) & -2.98 & (-3.06,\;-2.90)\\
        LCB & -2.62 & (-2.68,\;-2.56) & -2.77 & (-2.83,\;-2.7) & -2.32 & (-2.37,\;-2.27)\\
        CB & -2.62 & (-2.69,\;-2.55) & -2.69 & (-2.76,\;-2.63) & -2.39 & (-2.44,\;-2.34)\\
        RCB & -2.86 & (-2.93,\;-2.79) & -2.44 & (-2.5,\;-2.37) & -2.34 & (-2.39,\;-2.30)\\
        LWB & -2.33 & (-2.4,\;-2.26) & -2.61 & (-2.69,\;-2.53) & -2.27 & (-2.34,\;-2.20)\\
        LCMF & -2.51 & (-2.59,\;-2.43) & -2.50 & (-2.57,\;-2.42) & -2.07 & (-2.13,\;-2.01)\\
        DMF & -2.62 & (-2.68,\;-2.55) & -2.42 & (-2.49,\;-2.36) & -2.13 & (-2.18,\;-2.09)\\
        RCMF & -2.34 & (-2.41,\;-2.26) & -2.62 & (-2.7,\;-2.54) & -2.20 & (-2.26,\;-2.15)\\
        RWB & -2.48 & (-2.56,\;-2.4) & -2.29 & (-2.38,\;-2.19) & -2.01 & (-2.07,\;-1.95)\\
        SS &  -2.62 & (-2.71,\;-2.54) & -2.38 & (-2.46,\;-2.29) & -2.12 & (-2.21,\;-2.04)\\
        CF & -2.62 & (-2.71,\;-2.53) & -2.99 & (-3.09,\;-2.89) & -2.30 & (-2.38,\;-2.22)\\
        Winning &  -0.41 & (-0.47,\;-0.35) & -0.38 & (-0.45,\;-0.31) & -0.36 & (-0.4,\;-0.33)\\
        Losing & 0.06 & (0,\;0.11) & -0.01 & (-0.05,\;0.04) & -0.07 & (-0.12,\;-0.02)\\
        \hline
        \multicolumn{6}{l}{\bf Random effects:}\\
        Correlation & -0.36 & (-0.82,\;0.09) & -0.26 & (-0.75,\;0.23) & -0.05 & (-0.56,\;0.46)\\
        SD: success & 0.58 & (0.3,\;0.85) & 0.62 & (0.33,\;0.91) & 0.82 & (0.45,\;1.19)\\
        SD: pass & 0.50 & (0.27,\;0.73) & 0.24 & (0.12,\;0.36) & 0.46 & (0.25,\;0.67)\\
        \bottomrule
    \end{tabular}
    \label{tab:appRedInf}
\end{table}
\end{center}

\begin{center}
\begin{table}[!htbp]
\caption{\textcolor{blue}{Posterior summaries for Juventus Turin (with 4-4-2 formation) under Prior 1 described in Section \ref{sec:sensitivity}:
M refers to posterior medians and CI refers to 95\% posterior credible intervals.}}
\s
    \centering
    \footnotesize
    \makebox[\linewidth]{
        \begin{tabular}{lrc}
        & \multicolumn{2}{c}{\bf Juventus Turin} \\
          & M & CI\\
        \midrule
        \multicolumn{3}{l}{\bf Successful passes $\{S_{i_m} = 1\}$:}\\
        Intercept  & 3.38 & (2.96,\;3.80)\\
        Length of pass  & 0.00 & (-0.01,\;0.00)\\
        Forward pass & -0.62 & (-0.76,\;-0.48)\\
        Start: half & 0.25 & (0.09,\;0.41)\\
        End: third & -0.92 & (-1.07,\;-0.76)\\
        Air pass & -2.04 & (-2.18,\;-1.90)\\
        Winning & -0.05 & (-0.17,\;0.08)\\
        Losing & 0.04 & (-0.13,\;0.20)\\
        \hline
         \multicolumn{3}{l}{\bf Passes $\{i_m \to j_m\}$ given $\{S_{i_m} = 1\}$:}\\
        Graph distance & -0.70 & (-0.73,\;-0.67)\\
        Pass received  & 0.00 & (-8.12e-4,\;6.42e-4)\\
        \hline
        \multicolumn{3}{l}{\bf Holding times $h_m$:}\\
        GK & -2.80 & (-2.88,\;-2.73)\\
        LB & -2.08 & (-2.13,\;-2.03)\\
        LCB &  -2.38 & (-2.42,\;-2.33)\\
        RCB & -2.36 & (-2.41,\;-2.31)\\
        RB & -2.09 & (-2.14,\;-2.05)\\
        LW & -2.05 & (-2.11,\;-1.99)\\
        LCMF & -2.20 & (-2.25,\;-2.15)\\
        RCMF & -2.26 & (-2.3,\;-2.21)\\
        RW &  -2.16 & (-2.23,\;-2.1)\\
        SS & -1.80 & (-1.86,\;-1.74)\\
        CF & -1.93 & (-2,\;-1.86)\\
        Winning & -0.20 & (-0.23,\;-0.16)\\
        Losing & -0.15 & (-0.19,\;-0.10)\\
        \hline
        \multicolumn{3}{l}{\bf Random effects:}\\
        Correlation &  -0.30 & (-0.77,\;0.17)\\
        SD: success & 0.68 & (0.39,\;0.97)\\
        SD: pass &  0.43 & (0.24,\;0.62)\\
        \bottomrule
        \end{tabular}
        } 
    \label{tab:appRed_juventusInf}
\end{table}
\end{center}

\begin{center}
\begin{table}[!htbp]
\caption{\textcolor{blue}{Posterior summaries for Fiorentina, Crotone, and Inter Milan (with 3-5-2 formation) under Prior 3 described in Section \ref{sec:sensitivity}: 
M refers to posterior medians and CI refers to 95\% posterior credible intervals.}}
\s
    \centering
    \footnotesize
    \begin{tabular}{lrlrlrl}
        & \multicolumn{2}{c}{\bf Fiorentina} &  \multicolumn{2}{c}{\bf Crotone} &  \multicolumn{2}{c}{\bf Inter Milan} \s
        \\
          & M & CI & M & CI & M & CI\\
        \midrule
        \multicolumn{6}{l}{\bf Successful passes $\{S_{i_m} = 1\}$:}
        \\
        Intercept & 2.94 & (2.54,\;3.34) & 3.23 & (2.82,\;3.65) & 3.26 & (2.75,\;3.76)\\
        Length of pass  & 0.00 & (-0.01,\;0) & 0.00 & (-0.01,\;0) & 0.00 & (0,\;0.01)\\
        Forward pass & -0.56 & (-0.73,\;-0.39) & -0.88 & (-1.05,\;-0.7) & -0.84 & (-0.99,\;-0.69)\\
        Start: half & 0.17 & (-0.02,\;0.37) & -0.01 & (-0.21,\;0.18) & 0.30 & (0.12,\;0.47)\\
        End: third & -0.69 & (-0.87,\;-0.5) & -0.63 & (-0.82,\;-0.44) & -0.79 & (-0.97,\;-0.62)\\
        Air pass & -1.77 & (-1.93,\;-1.6) & -1.90 & (-2.06,\;-1.73) & -1.84 & (-1.99,\;-1.69)\\
        Winning & -0.13 & (-0.3,\;0.04) & -0.25 & (-0.45,\;-0.05) & -0.12 & (-0.25,\;0.02)\\
        Losing & -0.02 & (-0.18,\;0.14) & -0.12 & (-0.27,\;0.02) & 0.02 & (-0.16,\;0.21)\\
        \hline
        \multicolumn{6}{l}{\bf Passes $\{i_m \to j_m\}$ given $\{S_{i_m} = 1\}$:}\\
        Graph distance & -0.69 & (-0.73,\;-0.65) & -0.70 & (-0.74,\;-0.66) & -0.99 & (-1.02,\;-0.95)\\
        Pass received & 0.00 & (-2.0e-3,\;2.2e-3) & 0.00 & (-1.6e-3,\;3.8e-4) & 0.00 & (-1.5e-3,\;6.2e-05)\\
        \hline
        \multicolumn{6}{l}{\bf Holding times $h_m$:}\\
        GK & -3.23 & (-3.34,\;-3.13) & -2.86 & (-2.95,\;-2.77) & -2.98 & (-3.06,\;-2.9)\\
        LCB & -2.62 & (-2.69,\;-2.56) & -2.77 & (-2.83,\;-2.7) & -2.32 & (-2.37,\;-2.27)\\
        CB & -2.62 & (-2.69,\;-2.55) & -2.70 & (-2.76,\;-2.64) & -2.39 & (-2.44,\;-2.34)\\
        RCB & -2.86 & (-2.93,\;-2.79) & -2.44 & (-2.5,\;-2.37) & -2.34 & (-2.39,\;-2.3)\\
        LWB & -2.33 & (-2.4,\;-2.26) & -2.61 & (-2.69,\;-2.53) & -2.28 & (-2.34,\;-2.21)\\
        LCMF & -2.50 & (-2.58,\;-2.42) & -2.50 & (-2.57,\;-2.42) & -2.08 & (-2.13,\;-2.02)\\
        DMF & -2.62 & (-2.68,\;-2.55) & -2.42 & (-2.49,\;-2.36) & -2.13 & (-2.18,\;-2.09)\\
        RCMF & -2.34 & (-2.41,\;-2.26) & -2.62 & (-2.7,\;-2.55) & -2.20 & (-2.26,\;-2.15)\\
        RWB & -2.47 & (-2.55,\;-2.39) & -2.28 & (-2.37,\;-2.19) & -2.01 & (-2.07,\;-1.95)\\
        SS & -2.63 & (-2.72,\;-2.54) & -2.38 & (-2.46,\;-2.29) & -2.12 & (-2.2,\;-2.03)\\
        CF & -2.62 & (-2.72,\;-2.53) & -2.99 & (-3.09,\;-2.89) & -2.30 & (-2.38,\;-2.22)\\
        Winning & -0.42 & (-0.47,\;-0.36) & -0.38 & (-0.45,\;-0.3) & -0.36 & (-0.4,\;-0.33)\\
        Losing & 0.05 & (0,\;0.1) & -0.01 & (-0.05,\;0.04) & -0.08 & (-0.13,\;-0.02)\\
        \hline
        \multicolumn{6}{l}{\bf Random effects:}\\
        Correlation & -0.40 & (-0.87,\;0.07) & -0.32 & (-0.83,\;0.18) & -0.08 & (-0.61,\;0.45)\\
        SD: success & 0.58 & (0.32,\;0.84) & 0.60 & (0.34,\;0.86) & 0.78 & (0.45,\;1.11)\\
        SD: pass & 0.50 & (0.28,\;0.71) & 0.23 & (0.12,\;0.35) & 0.44 & (0.24,\;0.64)\\
        \bottomrule
    \end{tabular}
    \label{tab:appRedVague}
\end{table}
\end{center}

\begin{center}
\begin{table}[!htbp]
\caption{\textcolor{blue}{Posterior summaries for Juventus Turin (with 4-4-2 formation) using Prior 3 described in Section \ref{sec:sensitivity}:
M refers to posterior medians and CI refers to 95\% posterior credible intervals.}}
\s
    \centering
    \footnotesize
    \makebox[\linewidth]{
        \begin{tabular}{lrc}
            & \multicolumn{2}{c}{\bf Juventus Turin} \\
            & M & CI\\
            \midrule
            \multicolumn{3}{l}{\bf Successful passes $\{S_{i_m} = 1\}$:}\\
            Intercept & 3.35 & (2.93,\;3.77)\\
            Length of pass & 0.00 & (-0.01,\;0)\\
            Forward pass & -0.62 & (-0.75,\;-0.48)\\
            Start: half & 0.25 & (0.09,\;0.41)\\
            End: third & -0.92 & (-1.07,\;-0.76)\\\;
            Air pass & -2.05 & (-2.19,\;-1.91)\\
            Winning & -0.05 & (-0.18,\;0.08)\\
            Losing & 0.04 & (-0.13,\;0.2)\\
            \hline
            \multicolumn{3}{l}{\bf Passes $\{i_m \to j_m\}$ given $\{S_{i_m} = 1\}$:}\\
            Graph distance & -0.70 & (-0.73,\;-0.67)\\
            Pass received & 0.00 & (-7.7e-4,\;7.1e-4)\\
            \hline
            \multicolumn{3}{l}{\bf Holding times $h_m$:}\\
            GK & -2.80 & (-2.88,\;-2.73)\\
            LB & -2.08 & (-2.13,\;-2.03)\\
            LCB & -2.38 & (-2.42,\;-2.33)\\
            RCB & -2.36 & (-2.41,\;-2.31)\\
            RB & -2.09 & (-2.14,\;-2.05)\\
            LW & -2.06 & (-2.12,\;-2)\\
            LCMF & -2.20 & (-2.25,\;-2.15)\\
            RCMF & -2.26 & (-2.31,\;-2.21)\\
            RW & -2.17 & (-2.23,\;-2.11)\\
            SS & -1.81 & (-1.88,\;-1.74)\\
            CF & -1.94 & (-2.00,\;-1.87)\\
            Winning & -0.20 & (-0.23,\;-0.16)\\
            Losing & -0.15 & (-0.19,\;-0.1)\\
            \hline
            \multicolumn{3}{l}{\bf Random effects:}\\
            Correlation & -0.30 & (-0.77,\;0.18)\\
            SD: success & 0.66 & (0.37,\;0.94)\\
            SD: pass & 0.42 & (0.23,\;0.61)\\
            \bottomrule
        \end{tabular}
        } 
    \label{tab:appRed_juventusVague}
\end{table}
\end{center}

\clearpage

\section{Posterior predictive checks}
\label{sec:posterior.predictive.checks}

\begin{figure}[htb]
    \centering
    \includegraphics[width = 0.9\textwidth]{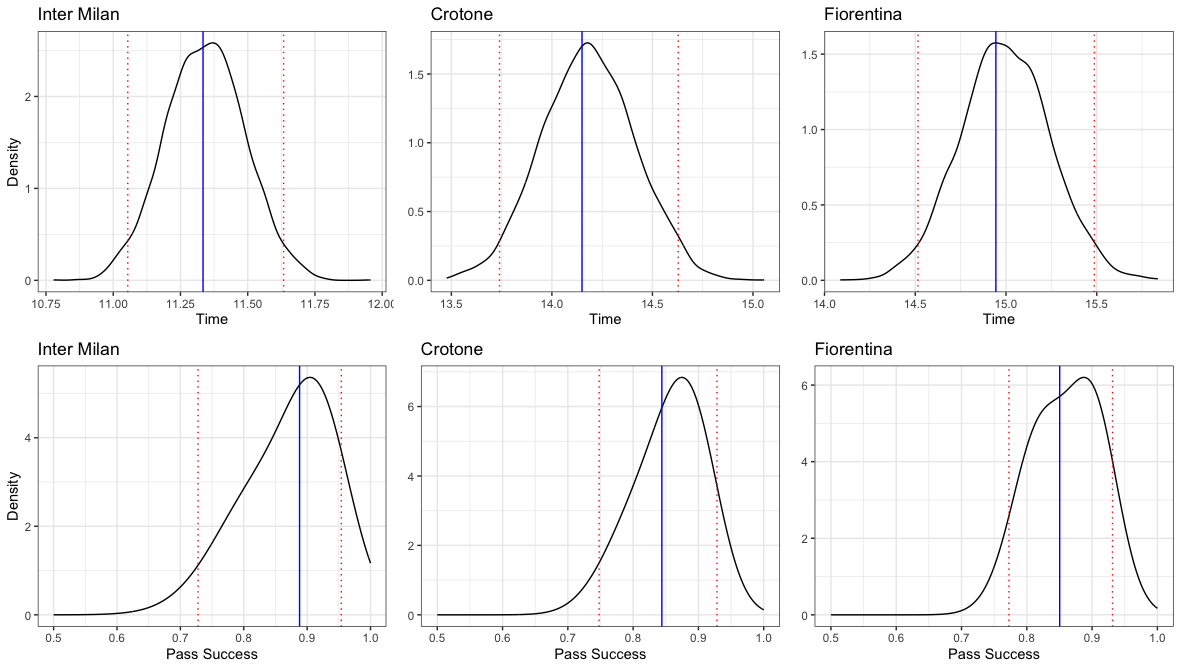}
    \caption{{Posterior predictions of the waiting times between passes and the proportions of successful passes by Inter Milan, Crotone, and Fiorentina during the 2020/21 season. The blue-colored solid vertical lines represent the mean of the observed waiting times and the observed proportions of successful passes, while the red-colored dotted vertical lines represent the 2.5\% and 97.5\% percentiles of the posterior predictions. 
    }}
    \label{fig:postpredcheck}
\end{figure}

\newpage

\section{Simulation results}
\label{sec:simulation.results}

\begin{figure}[htp]
\begin{center}
\includegraphics[scale=0.3]{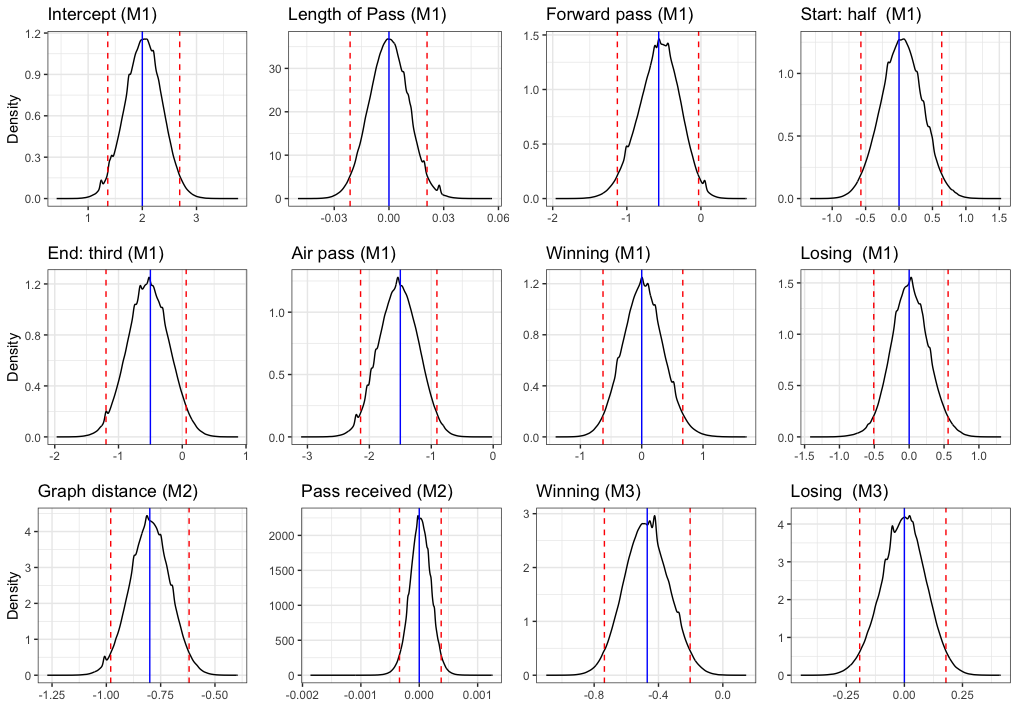}
\end{center}
\caption{{Simulation results:
    marginal posteriors of selected parameters based on 100 simulated soccer seasons,
    each with 1,000 passes.
    The blue-colored solid lines represent the data-generating parameters,
    while the red-colored dashed lines represent the 2.5\% and 97.5\% percentiles.
    M1, M2, and M3 refer to Module M1, M2, and M3 of the stochastic modeling framework specified in Section \ref{sec:app}, respectively.
    }}
    \label{ap:SimResults}
\end{figure}

\begin{center}
\begin{table}[!htbp]
\caption{\textcolor{blue}{Simulation results: 
data-generating parameters and posterior summaries of parameters based on one of the 100 simulated soccer seasons with 1,000 passes.
M is the median of the posterior means. 
CI shows the interval consisting of the 2.5\% and 97.5\% quantiles of the posterior means.}}
\s
    \centering
    \footnotesize
    \makebox[\linewidth]{
        \begin{tabular}{llrc}
            & \multicolumn{2}{c}{\bf Simulation} \\
            & Truth & M & CI\\
            \midrule
            \multicolumn{3}{l}{\bf Successful passes $\{S_{i_m} = 1\}$:}\\
            Intercept & 2.00 & 2.26 & (1.78,\;2.74) \\
            Length of pass & 0.00 &  0.00 & (-0.02,\;0.01)\\
            Forward pass & -0.57 &  -0.72 & (-1.1,\;-0.33)\\
            Start: half & 0.00 & 0.00 & (-0.43,\;0.43)\\
            End: third & -0.50 & -0.43 & (-0.88,\;0.02)\\
            Air pass &  -1.50 & -1.28 & (-1.72,\;-0.85)\\
            Winning & 0.00 & -0.11 & (-0.57,\;0.34)\\
            Losing & 0.00 & -0.14 & (-0.49,\;0.22)\\
            \hline
            \multicolumn{3}{l}{\bf Passes $\{i_m \to j_m\}$ given $\{S_{i_m} = 1\}$:}\\
            Graph distance & -0.80 & -0.83 & (-0.96,\;-0.69)\\
            Pass received &  0.00 & 0.00 & (-3.2e-4,\;2.1e-4)\\
            \hline
            \multicolumn{3}{l}{\bf Holding times $h_m$:}\\
            GK & -2.70 & -2.66 & (-2.94,\;-2.37)\\
            LCB & -2.70 & -2.84 & (-3.01,\;-2.66)\\
            CB & -2.70 & -2.66 & (-2.85,\;-2.48)\\
            RCB & -2.70 & -2.60 & (-2.8,\;-2.41)\\
            LWB & -2.70 & -2.71 & (-2.98,\;-2.44)\\
            LCMF & -2.70 & -2.74 & (-2.94,\;-2.55)\\
            DMF & -2.70 & -2.78 & (-2.96,\;-2.59)\\
            RCMF & -2.70 & -2.52 & (-2.72,\;-2.32)\\
            RWB &  -2.70 & -2.70 & (-2.92,\;-2.48)\\
            SS & -2.70 & -2.55 & (-2.83,\;-2.28)\\
            CF & -2.70 & -2.70 & (-2.97,\;-2.43)\\
            Winning & -0.47 & -0.58 & (-0.77,\;-0.39)\\
            Losing & 0.00 & 0.03 & (-0.12,\;0.17)\\
            \hline
            \multicolumn{3}{l}{\bf Random effects:}\\
            Correlation & 0.00 & -0.09 & (-0.88,\;0.71)\\
            SD: success & 0.00 & 0.20 & (0.01,\;0.38)\\
            SD: pass & 0.00 & 0.09 & (0,\;0.17)\\
            \bottomrule
        \end{tabular}
        } 
    \label{tab:appSim}
\end{table}
\end{center}

\end{appendix}


\end{document}